\documentclass[acmsmall,screen,nonacm]{acmart}

\usepackage{multirow}
\usepackage{hyperref}
\usepackage{enumitem} 
\usepackage{supertabular}
\usepackage{pdflscape}

\AtBeginDocument{%
  }

\setcopyright{acmlicensed}
\copyrightyear{2026}
\acmYear{2026}
\acmDOI{XXXXXXX.XXXXXXX}

\acmJournal{JACM}
\acmVolume{37}
\acmNumber{4}
\acmArticle{111}
\acmMonth{8}

\begin{document}

% \title{A Survey of Event Stream Modelling for Electronic Health Records}

\title{The Taxonomies, Training, and Applications of Event Stream Modelling for Electronic Health Records}
\author{Mingcheng Zhu}
\email{mingcheng.zhu@eng.ox.ac.uk}
\affiliation{%
  \institution{University of Oxford}
  \city{Oxford}
  \country{UK}
}

\author{Yu Liu}
\email{yu.liu@eng.ox.ac.uk}
\affiliation{%
  \institution{University of Oxford}
  \city{Oxford}
  \country{UK}
}

\author{Zhiyao Luo}
\email{zhiyao.luo@eng.ox.ac.uk}
\affiliation{%
  \institution{University of Oxford}
  \city{Oxford}
  \country{UK}
}

\author{Tingting Zhu}
\email{tingting.zhu@eng.ox.ac.uk}
\affiliation{%
  \institution{University of Oxford}
  \city{Oxford}
  \country{UK}
}


\begin{abstract}
The widespread adoption of electronic health records (EHRs) enables the acquisition of heterogeneous clinical data, spanning lab tests, vital signs, medications, and procedures, which offer transformative potential for artificial intelligence in healthcare. Although traditional modelling approaches have typically relied on multivariate time series, they often struggle to accommodate the inherent sparsity and irregularity of real-world clinical workflows. Consequently, research has shifted toward event stream representation, which treats patient records as continuous sequences, thereby preserving the precise temporal structure of the patient journey. However, the existing literature remains fragmented, characterised by inconsistent definitions, disparate modelling architectures, and varying training protocols. To address these gaps, this review establishes a unified definition of EHR event streams and introduces a novel taxonomy that categorises models based on their handling of event time, type, and value. We systematically review training strategies, ranging from supervised learning to self-supervised methods, and provide a comprehensive discussion of applications across clinical scenarios. Finally, we identify open critical challenges and future directions, with the aim of clarifying the current landscape and guiding the development of next-generation healthcare models.
\end{abstract}

\maketitle

\section{Introduction}

Electronic health records (EHRs) have become indispensable in modern healthcare, systematically capturing diverse and longitudinal patient information, including lab tests, vital signs, medications, diagnoses, and procedures~\cite{yang2023machine}. It serves as a fundamental data source for artificial intelligence (AI) in medicine, supporting a range of applications such as disease detection~\cite{placido2023deep}, clinical outcome prediction~\cite{rajkomar2018scalable, zhu2025bridging}, and personalised treatment recommendation~\cite{zhao2024leave, rudrapatna2025trial}. Until now, most EHR modelling frameworks have continued to rely on multivariate time series representations, which require dividing records into fixed temporal bins and imputing missing values~\cite {kazijevs2023deep, weerakody2021review, kazijevs2023deep}. Although such an approach facilitates the adoption of well-established time series models, it often fails to capture the inherent sparsity and irregular periodicity of real-world clinical data~\cite{xie2022deep}, resulting in significant information loss and the introduction of systematic bias~\cite {zhou2023missing}. 
To overcome these limitations, a new representation called event streams has been proposed~\cite{niu2024ehr}. Specifically, medical records are treated as a chronological sequence of atomic clinical events defined by three core components: the occurrence time, the event type (such as a diagnosis or medication), and the associated value (such as a lab result). This structure preserves exact details in patient records and reconstructs the original patient journey in chronological order~\cite{arnrich2024medical}. By inherently accommodating the irregularity and sparsity of clinical data, event streams bypass the requirement for division of records or imputation of missing values.

% Event stream modelling offers key advantages, including a more complete preservation of clinical information~\cite{estiri2020transitive}, a reduced risk of introducing bias~\cite{liu2022integrated}, and captures event relationships across the sequence that can be beneficial for downstream tasks~\cite{liu2020learning}.

Despite growing interest in event stream representation and EHR modelling, comprehensive reviews remain scarce. Existing surveys often suffer from notable scope limitations, particularly with respect to event stream definitions, modelling taxonomy, and clinical scenarios. As shown in Table~\ref{tab:review_comparison}, \citeauthor{shchur2021neural} \cite{shchur2021neural} provides an early overview, but focuses exclusively on point process methods and lacks a formal definition of event streams. \citeauthor{amirahmadi2023deep}~\cite{amirahmadi2023deep} defines event streams as sequences of hospital admissions, thereby overlooking finer-grained representations (e.g., hourly or event-level). Meanwhile, \citeauthor{zolyomi2025unified}~\cite{zolyomi2025unified} concentrates solely on self-supervised learning, neglecting a discussion of event stream modelling methods and the healthcare scenario. Consequently, the field lacks consensus on the fundamental definitions, modelling methods, training strategies, and clinical application scenarios, which impedes rigorous method comparison and practical adoption~\cite{amirahmadi2023deep}. 

To address the aforementioned gaps, this review systematically examines EHR event streams to consolidate fragmented knowledge and establish a consensus in this field. Specifically, Section~\ref{sec:search_strategy} details our literature search methodology. Section~\ref{sec:event_stream_definition} defines the EHR event stream, discusses its key characteristics, and compares it with traditional multivariate time series representations. Section~\ref{sec:method_taxonomy} introduces a taxonomy of modelling methods based on how they handle the time, type, and value of clinical events. Section~\ref{sec:training_strategy} discusses training strategies, ranging from supervised learning designed for the prediction of specific clinical outcomes to self-supervised paradigms that leverage the intrinsic structure of the clinical sequence to learn patient representations based on tasks. Section~\ref{sec:application} summarises clinical applications in various healthcare settings. Following this technical overview, Section~\ref{sec:challenges} discusses open research questions for the representation and modelling of event streams. Section~\ref{sec:future_directions} outlines promising future research directions. Finally, Section~\ref{sec:conclusion} concludes the paper with reflections on current progress and opportunities of the EHR event streams. A summary of the papers reviewed is shown in Appendix Table~\ref{append_table: Summary of papers}.

\begin{table}[h!]
  \centering
  \fontsize{7}{8.4}\selectfont
  \caption{Comparison of existing reviews on EHR event stream. \checkmark indicates coverage.}
  \vspace{-8px}
  \label{tab:review_comparison}
  \begin{tabular}{p{3.6cm} c c c c}
    \toprule
     &\citeauthor{shchur2021neural} (2021) & \citeauthor{amirahmadi2023deep} (2023) & \citeauthor{zolyomi2025unified} (2025) & \textbf{This Work} \\
    \midrule
    \textbf{Modelling Taxonomy} & & & & \\
    \quad Time Modelling & \checkmark &  & \checkmark & \checkmark \\
    \quad Type Modelling & \checkmark & \checkmark & \checkmark & \checkmark \\
    \quad Value Modelling &  & \checkmark & \checkmark & \checkmark \\
    \midrule
    \textbf{Training Strategy} & & & & \\
    \quad Supervised & \checkmark & \checkmark & & \checkmark \\
    \quad Self-Supervised &  & \checkmark & \checkmark & \checkmark \\
    \midrule
    \textbf{Healthcare Scenario} & & & & \\
    % \quad Healthcare Focused &  & \checkmark & & \checkmark \\
    \quad Intensive Care Unit (ICU) & \checkmark & \checkmark & \checkmark & \checkmark \\
    \quad Emergency Department (ED) &  &  &  & \checkmark \\
    \quad Primary Care &  &  & \checkmark & \checkmark \\
    \quad Hospital/Secondary Care &  &  &  & \checkmark \\
    \bottomrule
  \end{tabular}
\end{table}

\section{Searching Strategy}
\label{sec:search_strategy}

\begin{figure}[ht!]
    \centering
    \footnotesize
    \includegraphics[width=0.6\linewidth]{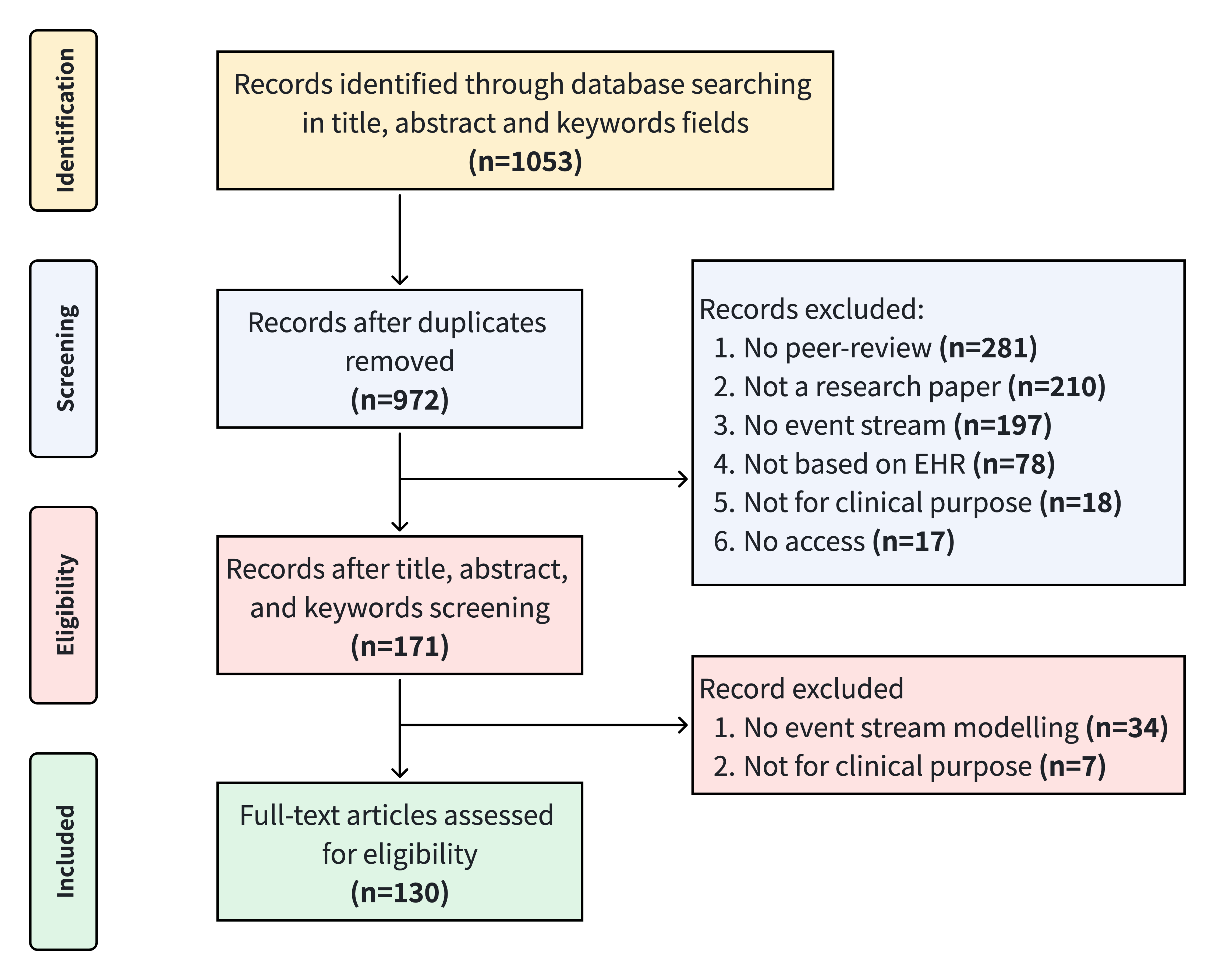}
    \caption{The flow diagram of the literature searching strategy.} 
    \label{fig:searching_strategy}
\end{figure}

The search process is illustrated in Figure~\ref{fig:searching_strategy}. Articles were eligible if they represented EHRs as event streams for deep learning modelling in healthcare applications. We searched titles, abstracts, and keywords in Scopus, PubMed, IEEE Xplore, ACM Digital Library, and Google Scholar (January 2026) using:
    \begin{quote}
    ("event stream*" OR "event sequence*" OR "clinical event*") AND ("deep learning" OR "machine learning" OR "Transformer" OR "large language model*" ) AND ("electronic health record*" OR "EHR" OR "electronic medical record*")
    \end{quote}
Our search strategy was limited to articles written in English and appearing in peer-reviewed journals or conference proceedings.

\section{Event Stream Representation of EHRs}
\label{sec:event_stream_definition}

\subsection{Definition of Event Stream Representation}

An EHR event stream for a patient is defined as a chronologically ordered sequence of clinical events. It can be represented formally as:
\begin{equation}
  \mathcal{X} = \{X_1, X_2, \dots, X_Q\},\quad \text{where}\quad X_q = \{x_{q,r} = (t_q, c_{q,r}, v_{q,r})\},\quad t_{1} < t_{2} < \dots < t_{Q},
  \label{eq:event_stream_definition}
\end{equation}
where $t_{i}$ is the timestamp for each event and $\mathbf{X}_q$, called observation, is an unordered set of events that occur simultaneously in $t_q$. Each observation contains a set of tuples where $c_{q,r} \in \mathcal{C}$ is the event type and $v_{q,r}$ is the corresponding value.
EHR event streams exhibit several properties relevant to modelling and analysis:
(1) \textbf{Scope granularity} -- each observation \(X_q\) represents a flexible granularity level, which can correspond to a single clinical event, a fixed time window, or a hospital visit, allowing for an adaptable analysis depending on the clinical context.
(2) \textbf{Temporal ordering} -- the increasing sequence of timestamps (\(t_1 < t_2 < \dots < t_Q\)) preserves the chronological order of clinical events.
(3) \textbf{Irregular timing} -- the timestamps \(t_q\) are not uniformly spaced, reflecting the inherent irregularity of clinical event occurrences in real-world settings.

\begin{figure}[h!]
    \centering
    \includegraphics[width=0.85\linewidth]{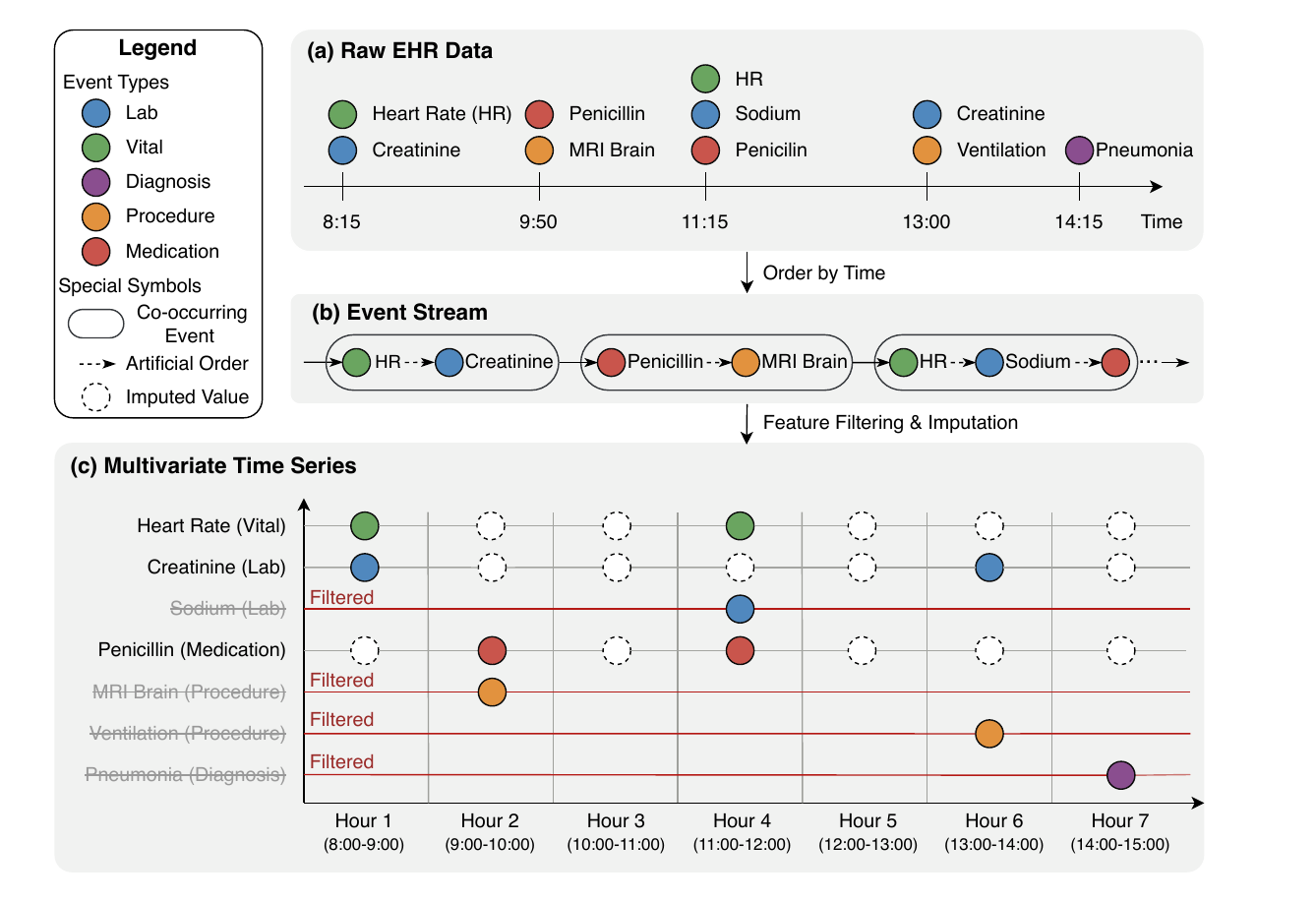}
    \vspace{-10px}
    \caption{Illustration of the event stream and multivariate time series representations of EHR. (a) Raw EHR Data: Clinical events spanning multiple heterogeneous types (e.g., vital signs, lab tests, medications, procedures, and diagnoses) are recorded at irregular, continuous timestamps. (b) Event Stream: All extracted clinical events are arranged chronologically into a sequence, with co-occurring events assigned an arbitrary relative order. (c) Multivariate Time Series: The data is converted into a structured grid by aligning variables to fixed time intervals (e.g., hourly windows). To reduce sparsity, features with a high missing rate are filtered out, and missing data for the retained variables are imputed, resulting in a dense, regularly sampled matrix.}
    \label{fig:ehrcuration_demo}
\end{figure}

\subsection{Advantages of Event Stream Representation}
The event stream representation preserves the original sequence and exact timing of all clinical events recorded in the EHRs. Clinical events are extracted and ordered chronologically. This approach provides a complete representation of the patient’s journey, as shown in Figure~\ref{fig:ehrcuration_demo}(b).  In multivariate time series that are used in most EHR modelling methods, a predefined set of clinical variables is selected, and patient records are discretised into regular time intervals. Each variable is aligned with these intervals, with multiple aggregated values and missing values imputed as necessary to create a dense representation (Figure~\ref{fig:ehrcuration_demo}(c)). 
Formally, let $\mathcal{T}^M = \{t_1^M, t_2^M, \dots, t_L^M\}$ denote the sequence of $L$ regularly spaced time points, and let $\mathcal{V}^M = \{v_1^M, v_2^M, \dots, v_D^M\}$ denote the set of $D$ selected clinical variables. The multivariate time series for a patient is then represented as:
\begin{equation}
    \mathbf{X}^M = [x_{l,d}^M],\  \mathbf{X}^M \in \mathbb{R}^{L \times D},
\end{equation}
where $x_{l,d}^M$ denotes the value of variable $v_d^M$ at time $t_l^M$.

The primary motivation for transitioning from multivariate time series to event stream representations is to mitigate biases introduced by manual preprocessing, particularly those associated with feature selection and imputation. By circumventing these steps, the curation of the event stream minimises the risk of discarding pertinent information and introducing artificial patterns, enabling more faithful and generalisable modelling~\cite{renc2024zero}. 

% Nevertheless, adopting the event stream representation results in extremely long and heterogeneous sequences~\cite{zucchet2023online}, necessitating advances in model architectures such as Transformers~\cite{vaswani2017attention} and self-supervised training strategies, including masked reconstruction~\cite{devlin2019bert} and next-token prediction~\cite{radford2019language}.

\begin{figure}[ht]
    \centering
    \includegraphics[width=0.65\linewidth]{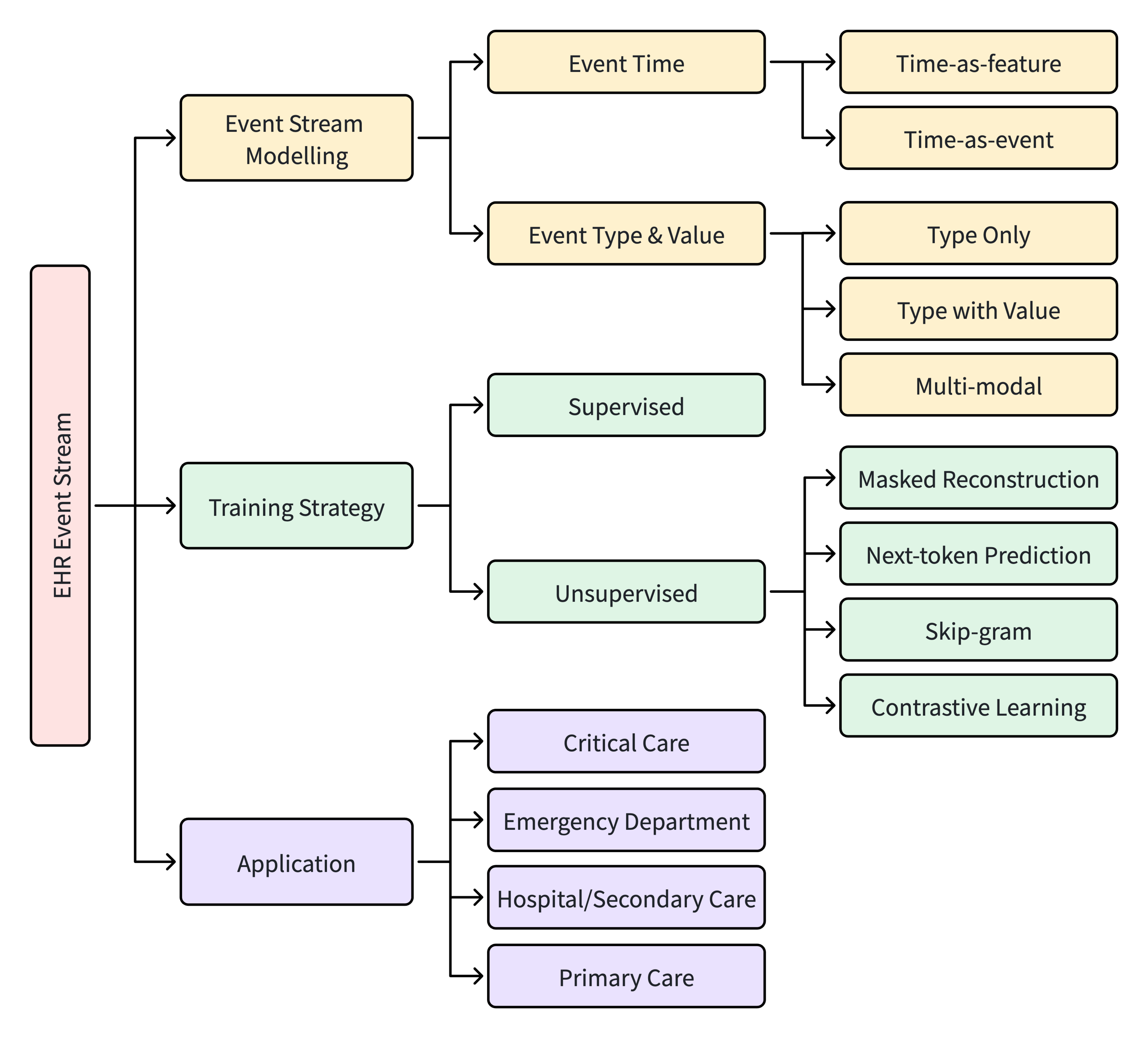}
    \vspace{-10px}
    \caption{Overview of event stream modelling. This overview categorises the field into three primary components, each highlighted using a distinct colour. Event stream modelling details the modelling methods for different components of an event. Training strategy outlines different learning paradigms for event stream models, including supervised and self-supervised methods. The application focuses on the different healthcare scenarios, including critical care, emergency department, hospital and secondary care, and primary care.}
    \label{fig:overview}
\end{figure}

\section{Event Stream Modelling Taxonomy}
\label{sec:method_taxonomy}
% This section introduces the event stream modelling methods. According to Equation~\ref{eq:event_stream_definition}, each event is defined as $x=(t,c,v)$. For each item, there are specified modelling methods.

This section establishes a systematic taxonomy for the modelling of EHR event streams. As defined in Equation~(\ref{eq:event_stream_definition}), each clinical event is represented by the tuple $x=(t,c,v)$, which encompasses its occurrence time $t$, type $c$, and associated value $v$. Rather than presenting isolated architectures, we examine the modelling strategies designed for each of these fundamental event components, categorised into event time modelling and the joint modelling of event types and values. It is important to note that a complete model must employ specific strategies for both the temporal dynamics and the heterogeneous event type and values.

\subsection{Event Time Modelling}

Effectively capturing the temporal dynamics of patient trajectories is a core challenge in EHR event stream modelling. Existing methodologies for integrating event timestamps can be broadly categorised into two foundational paradigms: time-as-feature and time-as-event.

\textbf{Time-as-feature\quad} In this approach, the timestamp $t$ is treated as a continuous input attribute associated with the type and value of the event. The fundamental objective is to transform this temporal information into a feature representation that can be integrated with clinical content to generate a unified event embedding. Methodologically, this is achieved through two mechanisms: explicit encoding, where dedicated modules extract temporal features, or implicit encoding, where temporal dynamics are captured by the sequential architecture of the model.

Explicit modelling strategies typically design specific embedding modules to represent event time. The most prevalent approach uses dedicated embedding layers within the Transformer architecture~\cite{vaswani2017attention}, where the encoded timestamp is concatenated with the event type and value vectors to form a unified input representation~\cite{shickel2022multi, lee2023learning, xu2023transehr}. Recently, rotary position embedding (RoPE)~\cite{su2024roformer} has been adopted to encode event timestamps, incorporating absolute time information into the rotation of event embeddings~\cite{song2025trajgpt, dai2025rotary}. RoPE enables the model to naturally capture relative temporal distances and extrapolate to future time horizons unseen during training, allowing models to learn from the temporal dynamics characteristic of longitudinal clinical histories. 
Beyond internal time encoding, some studies propose external modules for the integration of time and other features. For example, \citeauthor{wu2018modeling}~\cite{wu2018modeling} introduced a gating mechanism to fuse timestamps with event type and value prior to input into recurrent neural networks (RNNs)~\cite{schuster1997bidirectional}. This concept was further refined by subsequent studies~\cite{liu2018learning, lee2020multi, liu2021multi}, which designed RNN cells to explicitly modulate the information flow based on the temporal intervals between events. Additionally, some methods employ tokenisation, where timestamps are treated as discrete tokens and processed via Word2Vec~\cite{mikolov2013efficient} to generate semantic embeddings that inherently capture temporal context~\cite{lu2020clinical, lin2020medical, rethmeier2021efficare}.

In contrast to methods that rely on explicit time features, other architectures model temporal dynamics implicitly by leveraging the sequential order of the event stream. For Transformer models, this is achieved through positional encoding mechanisms, which assign a unique vector to each step to indicate its relative position within the sequence rather than its absolute timestamp~\cite{shou2023concurrent, liu2020learning}. Similarly, RNNs encode temporal progression implicitly through internal memory states that update recursively as the model traverses the sequence~\cite{nigo2024deep, wu2023iterative, barren2021improving}. Furthermore, \citeauthor{coombes2022sequences}~\cite{coombes2022sequences} adapt genomics methodologies, generating embeddings by segmenting the sequence into overlapping substrings and using their frequency distributions to represent temporal patterns. 
A critical limitation of the implicit approaches is the co-occurring events as defined in Equation (\ref{eq:event_stream_definition}). Flattening $\mathbf{X}_q$ into a linear sequence introduces artificial temporal dependencies, misleading the model by inferring causal relationships between concurrent events where none exist. To avoid imposing an arbitrary order, recent studies proposed graph-based representations. Specifically, the authors in~\cite{yang2024dynamic, zhu2022temporal, jiang2020learning} construct graphs where co-occurring events serve as nodes, allowing relationships to be captured in an order-agnostic manner. Features extracted by a graph neural network (GNN) are subsequently fed into a Transformer to model the temporal evolution of these graph states. In a different approach, \citeauthor{xie2024hgthp}~\cite{xie2024hgthp} integrates all events into a single global graph, utilising edges to model co-occurrence relationships across distinct timestamps.

\textbf{Time-as-event\quad} Unlike time-as-feature, the timestamp is treated as a distinct event within the sequence. Here, a special event type of time is introduced, typically represented as $(c_\text{time}, t)$, and is placed before the set of clinical events that occur at time $t$. The time-as-event sequence $\mathcal{X}^\text{tae}$ is thus structured as a sequence of event sets:

\begin{equation}
\mathcal{X}^\text{tae} = \left(\mathbf{X}^\text{tae}_q\right)_{q=1}^Q, \quad 
\text{where} \quad \mathbf{X}^\text{tae}_q = \left\{ (c_\text{time}, t_q), (c_{q,1}, v_{q,1}), \ldots, (c_{q,m_q}, v_{q,m_q}) \right\},
\end{equation}
where $\mathbf{X}^\text{tae}_q$ is the $q$-th observation that occurs at time $t_q$, and $m_q$ is the number of clinical events within that set. Although this formulation appears structurally distinct from the definition in Equation~\ref{eq:event_stream_definition}, where time is an intrinsic attribute of every tuple, the two representations are fundamentally equivalent. The Time-as-event format merely factorises the shared timestamp. The original event set $\mathbf{X}_q$ can be reconstructed by placing the timestamp $t_q$ of the special time token on each associated clinical pair $(c_{q,r}, v_{q,r})$. 
Specifically, \citeauthor{renc2024zero}~\cite{renc2024zero} designed a vocabulary with special time interval tokens, representing durations ranging from five minutes to over six months. These tokens are inserted at the beginning of each co-occurring event set to explicitly demarcate temporal boundaries. The entire sequence is then processed by a GPT-2 architecture~\cite{radford2019language} using an autoregressive objective, enabling the model to dynamically determine whether to generate a subsequent clinical event within the current timestamp or a time token signifying a transition to the next observation. In a text-centric approach, authors in ~\cite{wu2024instruction, zhu2025medtpe, chen2026cross} serialise the timestamp, event types, and values into a unified natural-language text, with the timestamp serving as the initial context. This composite text is subsequently processed by a large language model (LLM), allowing the temporal dynamics to be interpreted through the model's semantic understanding. 

\textbf{Summary\quad} 
The fundamental distinction between time-as-feature and time-as-event lies in the modelling of a shared timestamp $t_q$ within an observation $\mathbf{X}_q$. The \textit{time-as-feature} method integrates $t_q$ directly into every atomic event tuple, ensuring that each event representation preserves its full temporal context independently. However, this results in representational redundancy, as the identical timestamp is encoded repeatedly for all co-occurring events. In contrast, the \textit{time-as-event} approach factorises the observation by representing $t_q$ as a single distinct token that precedes the set of associated clinical events. Although this eliminates redundancy, it creates a contextual dependency. The temporal grounding of a clinical event is no longer intrinsic to its individual embedding, but must be inferred from the preceding time token via the model's sequential processing mechanism. 
The choice between these two approaches should be determined by the modelling objective. When the goal is to extract meaningful event embeddings, the \textit{time-as-feature} representation is generally more appropriate. It assigns to each event an explicit temporal attribute, allowing the model to jointly learn the relationship between time, event types, and the corresponding values~\cite{xu2023transehr}. In contrast, for generative modelling or sequence forecasting tasks, the \textit{time-as-event} representation is often more effective. By treating time as a distinct event that precedes a group of co-occurring events, this approach enables the model to predict not only what events will occur next but also when they will occur, facilitating the joint generation of temporal and clinical dynamics~\cite{renc2024zero}.

\subsection{Event Type and Value Modelling}
The modelling of the type of event $c$ and the value $v$ is inherently coupled, as the semantic meaning of an event value is linked to its associated type. A fundamental challenge in this domain is the heterogeneity of clinical data. The vast combination of event types and values results in a representation space that is both high-dimensional and sparse. To address this complexity, existing strategies can be classified into three distinct categories: type-only modelling, type with value modelling, and multimodal modelling. These categories represent a trend of increasing heterogeneity, advancing from simplified abstractions that consider only categorical types to complex frameworks that integrate continuous values and unstructured modalities.

\textbf{Type-only Modelling\quad}

In this approach, only the type of event \(c\) is modelled, while the value of event \(v\) is ignored. Each event is represented solely by its type, simplifying the modelling process. Formally, the event stream is represented as:
\begin{equation} 
\mathcal{X}^\text{to} = \{\mathbf{X}^\text{to}_q\}_{q=1}^{Q}, \quad \mathbf{X}^\text{to}_q = \left\{\mathbf{x}^\text{to}_{q,r}=(t_{q}, c_{q,r})\right\}_{r=1}^{|\mathbf{X}^\text{to}_q|}, \quad t_{1} < t_{2} < \dots < t_{Q}. 
\end{equation} 
A prominent approach models the type-only event stream using the temporal point process (TPP)~\cite{shchur2021neural}. The TPP models explicitly quantify the temporal dependencies between the types of clinical events within the continuous-time formalism~\cite{diggle1994parameter}. The central component is the conditional intensity function, which defines the instantaneous probability of an event of type $c$ occurring at time $t_i$, conditioned on historical observations $\mathcal{H}_{t_i} = \{ \mathbf{X}_q \mid t_q < t_i \}$. Formally, this is defined as: 
\begin{equation} 
\lambda_c(t_i \mid \mathcal{H}_{t_i}) = \mu_c + \sum_{q=1}^{ |\mathcal{H}_{t_i}|}\sum_{r=1}^{|\mathbf{X}_{q}|} \alpha_{q, r, c} , g(t_i - t_{q}), 
\end{equation}
where $\mu_c$ represents the baseline intensity of type $c$, and the nested summation aggregates the influence of all historical events in $\mathcal{H}_{t_i}$. The parameter $\alpha_{q, r, c}$ measures the excitation effect that a previous event type $c_{q, r}$ exerts on the current type $c$, while $g(\cdot)$ denotes a temporal kernel function describing how this influence decays over time~\cite{gao2021causal}. 
Classical implementations, such as the Hawkes process~\cite{shchur2021neural}, utilise this formulation to predict the precise timing of target events~\cite{weiss2017piecewise, gao2021causal}. More recently, hybrid architectures have emerged that integrate deep learning with TPPs. These studies typically employ neural networks to extract dense embeddings for time and type, which are subsequently fed into the TPP intensity function to enhance predictive performance~\cite{ru2022sparse, xie2024hgthp}. Similarly, \citeauthor{song2022tri}~\cite{song2022tri} developed a Transformer-based model to encode the sequence of past observations $\mathcal{H}_{t_i}$, using the resulting contextual representation to parameterise the intensity function for future event prediction. 

Another approach leverages a Transformer to process the event stream as a flat sequence of distinct input units. In this framework, each token represents a composite embedding that integrates both the event timestamp and its clinical type~\cite{zhang2022adadiag, niu2024ehr}. To preserve the hierarchical nature of patient history, these models often introduce a dedicated "visit segment" embedding. This feature serves as an explicit boundary marker, allowing the model to distinguish between separate clinical encounters while processing the continuous stream. Alternatively, hierarchical architectures~\cite{qiao2018pairwise, bai2018interpretable} have been developed to explicitly model these distinct temporal scales. These systems typically employ a dual-level mechanism: a lower-level RNN captures local, short-range dependencies among events within a single visit, while a higher-level additive attention mechanism~\cite{bahdanau2014neural} aggregates these visit representations to model long-range clinical evolution throughout the patient's entire history. 

A third approach quantifies event occurrences within an observation $\mathbf{X}_q$, using these counts as features. For example, the authors~\cite{coombes2022sequences, jiang2020learning} transform event streams into event count sequences, which are then used to measure patient similarity. Specifically, \citeauthor{wang2024twin}~\cite{wang2024twin} first aggregated administered medications and adverse drug effects into visit-level count vectors. This sequence is then fed into an auto-regressive Transformer~\cite{radford2019language} to predict the conditional probability of events in the next visit, enabling the generation of synthetic patient trajectories for digital twin applications.

\textbf{Type and Value Modelling\quad}

This approach jointly models the type of event $c$ and its corresponding value $v$. A central challenge in this domain is the heterogeneity of clinical values, which encompasses both categorical values (i.e., demographics or medical intervention) and continuous numerical values (i.e., vital signs). Existing methodologies are primarily distinguished by their treatment of continuous data, which is typically managed through either discretisation into categorical values or encoding of continuous values. 
One approach discretises continuous event values into categorical bins. For example, the authors in~\cite{renc2024zero, renc2025foundation} convert continuous values into quantile-based bins. Each bin is then treated as an independent token that follows the main event type token. This modelling paradigm unifies the format for both categorical and continuous events and separates the value from the type, thereby reducing model complexity. Similarly, Lin et al.~\cite{lin2022mr} combine descriptive and continuous values, segmenting them into distinct chunks, each of which is treated as an independent token for Transformer-based prediction. 
Alternative approaches preserve the full resolution of continuous values by encoding them directly. The authors in~\cite{oufattole2024event, wu2023iterative, theodorou2025improving} use a dedicated numerical encoder to produce a value embedding, which is subsequently concatenated with the event type and timestamp representations to form a comprehensive vector. Building upon this framework, \citeauthor{xu2023transehr}~\cite{xu2023transehr} developed a triplet encoding mechanism composed of three parallel processing heads. This architecture simultaneously employs a discrete embedding layer for categorical types, a linear projection for continuous values, and a trigonometric encoding module for temporal information. These heterogeneous features are then fused to generate a unified event embedding, which is processed by a Transformer to model the sequence.

\textbf{Multimodal Modelling\quad}
Conceptually, multimodal modelling can be viewed as an advanced extension of type and value modelling. While standard value modelling handles low-dimensional numerical or categorical results, the event value $v$ can also encompass highly complex, unstructured data associated with specific event types $c$. For example, a radiology event yields a high-dimensional medical image (e.g., X-rays, CT scans), and a consultation event generates free-text clinical notes or bio-signals. Multimodal modelling has its own distinct category due to the architectural necessity of employing modality-specific encoders to process these heterogeneous values before they can be integrated into the chronological event stream.

Current multimodal modelling for EHR event streams focuses predominantly on integrating structured EHR data with unstructured clinical text. The most straightforward strategy is direct concatenation, where separate embedding vectors generated for the event stream and clinical notes are joined to form a single, contiguous input representation~\cite{lin2022mr, he2020attention}. Similarly, \citeauthor{li2022next}~\cite{li2022next} implement early-stage fusion by projecting multimodal features within each day into unified embeddings prior to sequence processing. Alternatively, taking the event stream from the EHR and clinical text as two parallel sources, \citeauthor{lu2022multi}~\cite{lu2022multi} employs cross-attention mechanisms, with clinical events serving as active queries that selectively attend to the relevant sections of the unstructured clinical notes. This architecture allows the model to fuse information based on semantic relevance. \citeauthor{ding2024distilling}~\cite{ding2024distilling} propose a knowledge distillation method that uses LLMs to process clinical text and subsequently transfer the resulting semantic embeddings to train the event stream model. 
Moving beyond EHR data and clinical text, \citeauthor{lee2023learning}~\cite{lee2023learning} addresses the challenge of high-dimensional heterogeneity by combining EHR data, X-ray images, and text. Their method projects all modalities into the shared latent space with linear projection and utilises special tokens for different modalities. These special tokens act as bridges, compressing and routing information across the different modality-specific encoders to generate the patient representation.

\textbf{Summary\quad} 
The strategies mentioned above highlight a fundamental trade-off between architectural complexity and semantic granularity and the modalities of the representation. Type-only modelling prioritises computational simplicity by abstracting events into categorical tokens, thereby discarding specific measurement details. In contrast, type and value modelling enhance informational fidelity by integrating numerical measurements. Multimodal modelling represents the highest degree of complexity, necessitating specialised encoders to process unstructured data, such as clinical notes and medical imaging, and the module of modality integration. 
A discernible trend in the field is the progression from simplified type-only abstractions toward comprehensive multimodal modelling methods. This evolution is propelled by two primary drivers: the availability of large-scale, multi-source EHR datasets and the evolution of deep learning architectures. Modern architectures, particularly Transformers and pre-trained LLMs, now possess the capacity to effectively synthesise high-dimensional, multimodal inputs, enabling the construction of holistic patient representations that were previously computationally infeasible.

\section{Training Strategies for EHR Event Stream Models}
\label{sec:training_strategy}

Training strategies for EHR event stream models can be broadly categorised into two main paradigms: supervised learning and self-supervised learning. Supervised learning involves training models on labelled datasets, where each event stream is associated with specific outcomes or labels. Self-supervised learning takes advantage of the inherent structure of the event streams to create the pretraining task, enabling models to learn useful representations without explicit labels. 

\subsection{Supervised Learning}
In supervised learning, a deep learning model is trained to map input event streams to downstream tasks. Formally, given a dataset of $M$ patients, where each patient $j$ is represented by an event stream $\mathcal{X}_j$ and an associated label $y_j$, the model is trained to minimise a loss function. The overall supervised loss is defined as:
\begin{equation}
    \mathcal{L}_\text{supervised} = \sum_{j=1}^{M} \ell\bigl(f_\theta(\mathcal{X}_j), y_j\bigr),
\end{equation}
where $f_\theta$ is the deep learning model for event stream modelling, and $\ell(\cdot, \cdot)$ is a task-specific loss function. Specifically, the authors in~\cite{li2020marrying, mei2024collaborative, dong2020machine} formulate future event prediction as a classification task and train the model using the cross-entropy loss. In contrast, other works~\cite{qiao2018pairwise, hansen2023patient, chowdhury2024stratifying} focus on continuous targets, formulating the prediction of future measurement values as regression tasks using mean squared error (MSE) or root mean squared error (RMSE), and similarly treating time-to-event prediction as a regression in the time interval $\Delta t$. Furthermore, the authors in~\cite{liu2021multi, barren2021improving} adopt a multi-task learning strategy, training a shared event stream encoder with distinct prediction heads to simultaneously optimise a weighted sum of the individual task losses.

\begin{figure}[htp]
    \centering
    \includegraphics[width=0.93\linewidth]{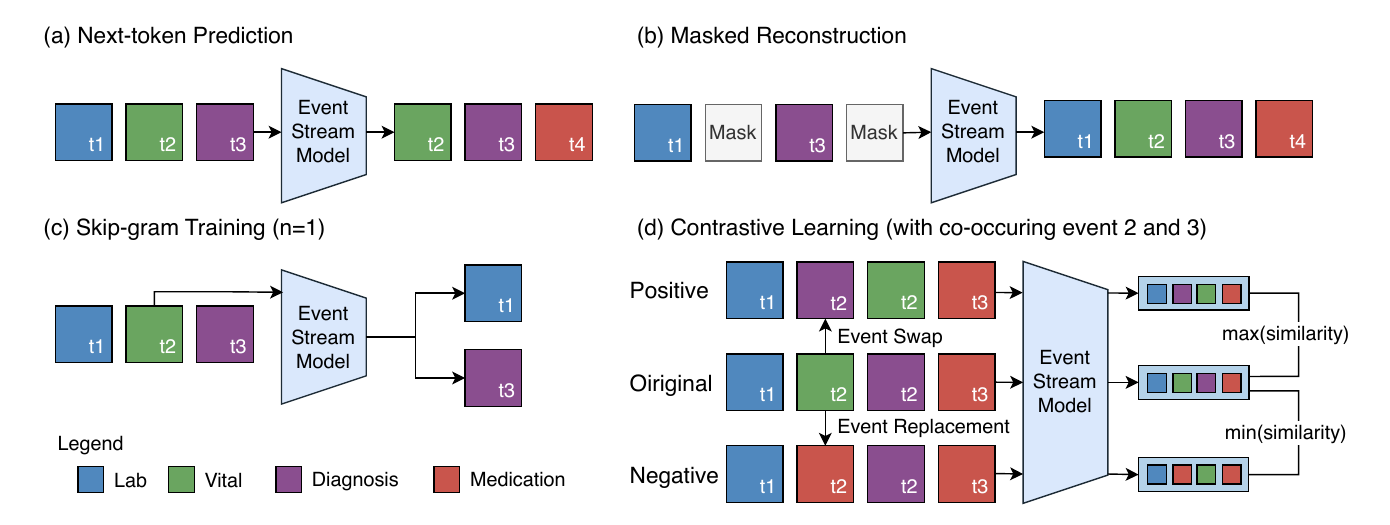}
    \vspace{-5px}
    \caption{Illustration of self-supervised learning methods for EHR event stream modelling. The four main approaches are: (a) Next-token prediction, where the model learns to predict the subsequent event in the sequence given the preceding events. (b) Masked reconstruction, where events in the input sequence are deliberately hidden (masked), and the model is trained to reconstruct the complete sequence. (c) Skip-gram ($n=1$), where the model uses a single target event (e.g., $t_2$) to predict its surrounding context events (e.g., $t_1$ and $t_3$). (d) Contrastive learning, which trains the model to maximise the embedding similarity between an original sequence and a positive pair that are generated via event swapping of co-occurring events, while minimising the similarity against a negative pair that is generated via event replacement.}
    \label{fig:ssl_methods}
\end{figure}

\subsection{Self-supervised Learning}

Self-supervised can be further categorised into four main approaches, as illustrated in Figure~\ref{fig:ssl_methods}: \textit{next-token prediction}, \textit{masked reconstruction}, \textit{skip-gram training}, and \textit{contrastive learning}. Each approach leverages the intrinsic structure of the event stream to create pretraining tasks that enable models to learn meaningful representations without relying on explicit labels.

\textbf{Next-token Prediction\quad}
Next-token prediction trains a model to predict the subsequent event in a sequence given all preceding events~\cite{radford2019language}, as illustrated in Figure~\ref{fig:ssl_methods}(b). This approach encourages the model to learn the temporal dependencies and sequential patterns within the event stream. Formally, the model is trained to minimise the following prediction loss:
\begin{equation}
    \mathcal{L}_\text{next} = \sum_{i=1}^{N-1} \ell\bigl(\hat{\mathbf{x}}_{i+1}, \mathbf{x}_{i+1}\bigr),
\end{equation}
where $\hat{\mathbf{x}}_{i+1}$ is the prediction of the next event by the model, conditioned on the history of events up to $\mathbf{x}_i$. 
Implementation strategies vary according to the granularity of the prediction target. The authors in~\cite{tariq2025adaptable, lu2023multi, yang2023transformehr} formulate the task as a prediction of the next-visit, where the objective is to foresee the entire set of events in the subsequent observation $\mathbf{X}_{q+1}$, conditioned on the sequence of previous visits completed $\{\mathbf{X}_1, \dots, \mathbf{X}_q\}$. Alternatively, other methods~\cite{guo2024multi, yang2023transformehr} operate at the atomic event level, training the model to predict the specific next item $x_{q,r+1}$ within an observation $\mathbf{X}_{q}$. In these methods, the conditioning context comprises the history of prior observations $\{\mathbf{X}_1, \dots, \mathbf{X}_{q-1}\}$ augmented by the local sequence of events $\{x_{q,1}, \dots, x_{q,r}\}$ already processed within the current observation $\mathbf{X}_q$.  
Other works adapt event embeddings for use with pretrained LLMs via next-token prediction. For example, \citeauthor{wu2024instruction}~\cite{wu2024instruction} embeds each event using ClinicalBERT~\cite{huang2019clinicalbert} and adapts these event embeddings for pretrained LLMs using a linear projection layer. The linear projection layer is trained to predict the next event in the sequence. Similarly, \citeauthor{ding2024distilling}~\cite{ding2024distilling} develops a specific encoder for events and trains this encoder to align with the hidden states of a pretrained LLM using next-token prediction. In addition, \citeauthor{zhu2025medtpe}~\cite{zhu2025medtpe} designs a specific tokenisation method for clinical events and integrates these tokens into the embedding space of a pretrained LLM with next-token prediction training.

\textbf{Masked Reconstruction\quad}
The masked reconstruction approach involves masking certain parts of the input event stream and training the model to reconstruct the original unmasked data~\cite{devlin2019bert}, as illustrated in Figure~\ref{fig:ssl_methods}(a). By learning to predict this missing information, the model captures the underlying structure of the event stream and the relationships among events. Formally, given an event stream, a subset of its events is randomly masked to create a corrupted version, $\tilde{\mathcal{X}}$. The model is then trained to minimise the reconstruction loss:

\begin{equation}
    \mathcal{L}_\text{recon} = \sum_{q=1}^{Q} \sum_{r=1}^{|\mathbf{X}_q|} \mathbb{I}_{\text{masked}}(q,r) \cdot \ell\bigl(\hat{\mathbf{x}}_{q,r}, \mathbf{x}_{q,r}\bigr),
\end{equation}
where $\mathbb{I}_{\text{masked}}(q,r)$ is an indicator function that equals 1 if event $\mathbf{x}_{q,r}$ is masked, $\hat{\mathbf{x}}_{q,r}$ is the model's reconstruction of that masked event, and $\ell(\cdot, \cdot)$ is a suitable loss function (e.g., cross-entropy for categorical events or MSE for continuous values). 
Specifically, the authors in~\cite{pang2021cehr, niu2024ehr, karami2023point, karami2024tee4ehr} apply masked reconstruction to event streams by randomly masking the event time and type within each observation $\mathbf{X}_q$. The model is trained to reconstruct these masked events based on the surrounding context. Another approach~\cite{xu2023transehr} further masks the event values, training the model to reconstruct the event time and type with a cross-entropy loss and the event value with an MSE loss. Other approaches~\cite{zhang2022adadiag, lin2022mr} combine masked reconstruction with adversarial training to mitigate the distribution shift among patient groups from different hospitals or with different admission times.

\textbf{Skip-gram Training\quad}
Skip-gram training draws inspiration from natural language processing, where the model learns to predict surrounding words given a target word~\cite{mikolov2013efficient}, as illustrated in Figure\ref{fig:ssl_methods}(c). In the context of EHR event streams, the model is trained to predict neighbouring events based on a target event, thereby capturing contextual relationships within the event stream. Formally, for each event $\mathbf{x}_i$ in the sequence, the model is trained to minimise the following loss:
\begin{equation}
    \mathcal{L}_\text{skip-gram} = - \sum_j^{N^\text{nei}(i)} \log P(\mathbf{x}_j \mid \mathbf{x}_i),
\end{equation}
where $N^\text{nei}(i)$ denotes the set of neighbouring events within a predefined context window around event $\mathbf{x}_i$, and $P(\mathbf{x}_j \mid \mathbf{x}_i)$ is the probability of event $\mathbf{x}_j$ given the target event $\mathbf{x}_i$. 
Implementation of this strategy typically focuses on learning dense vector representations for clinical concepts. Specifically, multiple studies~\cite{henriksson2016ensembles, xiang2019time, lin2020medical} adopt the event-as-token paradigm to generate event embeddings. By optimising the prediction of co-occurring or temporally adjacent events within a specified window, these models effectively capture the semantic and sequential associations between different clinical codes in a data-driven manner.

\textbf{Contrastive Learning\quad}
Contrastive learning functions by generating alternative views of the event stream and training the model to pull semantically similar (positive) pairs together while pushing dissimilar (negative) pairs apart~\cite{hu2024comprehensive}. Formally, for a given anchor event stream $\mathcal{X}$, the framework generates a positive sample $\mathcal{X}^+$ (typically via data augmentation) and identifies a set of $K$ negative samples $\{\mathcal{X}^-_k\}_{k=1}^K$, which are often drawn from other sequences in the same training batch. The model is optimised to minimise the contrastive loss, defined as:
\begin{equation}
\mathcal{L}_\text{contrastive} = - \log \frac{e^{\text{sim}(f_\theta(\mathcal{X}), f_\theta(\mathcal{X}^+)) }}{e^{\text{sim}(f_\theta(\mathcal{X}), f_\theta(\mathcal{X}^+))} + \sum_{k=1}^{K} e^{\text{sim}(f_\theta(\mathcal{X}), f_\theta(\mathcal{X}^-_k))}},
\end{equation}
where $f\theta(\cdot)$ denotes the encoder network, $\text{sim}(\cdot, \cdot)$ is a similarity metric (such as cosine similarity). This objective ensures that the model learns a representation space in which the anchor is aligned closer to its positive variant than to any of the $K$ negative distinct sequences. 
Contrastive learning methods vary primarily in how they construct positive pairs. One common strategy is data augmentation. The authors in~\cite{li2022hi, zhang2020diagnostic} generate positive samples by randomly cropping or masking parts of the original event stream. Specifically, a positive view is created by masking individual events or dropping entire observations, while negative samples are drawn from the event streams of unrelated individuals. This forces the model to recognise the patient's trajectory despite missing information. 
A second strategy exploits the inherent permutation invariance of co-occurring events. \citeauthor{ma2024memorize}~\cite{ma2024memorize} construct positive samples randomly by shuffling the order of events within a single observation $\mathbf{X}_q$. Since the events within an observation occur concurrently, their specific serialisation in the data stream is an artefact of recording rather than a meaningful temporal sequence. Consequently, this approach trains the model to generate consistent representations invariant to these arbitrary permutations, ensuring that the embedding captures the composition of the event set rather than an artificial order. 
A third approach leverages temporal consistency within the patient's history. The authors in~\cite{zhang2022forecasting, oufattole2024event} divide the event stream into sub-sequences based on time windows or visit boundaries. In this setup, the model maximises the similarity between two different sub-sequences belonging to the same patient (positive pair) while minimising the similarity with sub-sequences from different patients (negative pair). This encourages the model to learn patient-specific features that remain stable over different time points.

\textbf{Summary\quad} 
Although all self-supervised learning strategies share the goal of leveraging unlabelled data, they differ in their training objectives and downstream utility. Next-token prediction focuses on modelling sequential dependencies and temporal evolution. These methods are particularly effective for forecasting tasks (e.g., predicting future diagnoses) or when using autoregressive architectures such as Transformers that require a deep understanding of the trajectory of the patient~\cite{rasmy2021med, renc2024zero}. 
In contrast, associative approaches like masked reconstruction and skip-gram training prioritise local semantic relationships. By treating events as words in a context window, this method excels at capturing the co-occurrence similarities between clinical events, making it ideal for generating static code embeddings or representations where long-range temporal structure is less critical~\cite{henriksson2016ensembles}. 
Finally, contrastive learning focuses on separation and invariance. By explicitly distinguishing between positive and negative sample pairs, this strategy is well-suited for tasks requiring robust whole-sequence representations, such as patient similarity search or classification under noisy conditions where permutation invariance for co-occurring events is beneficial~\cite{amirahmadi2025trajectory}.  
In summary, the choice of strategy depends on the specific modelling requirements: next-token prediction is superior for temporal dynamics, masked reconstruction and skip-gram are good for semantic embedding, and contrastive learning for robust global representation.

\section{Healthcare Applications of EHR Event Stream}
\label{sec:application}

Event stream modelling has been adopted across a wide range of healthcare settings, addressing both acute and chronic care needs. Different scenarios present unique characteristics of event streams and have different interests in terms of clinical tasks and outcomes. 

\textbf{Critical Care\quad} Critical care, particularly in the intensive care unit (ICU), treats patients with life-threatening acute conditions that require continuous monitoring and intervention~\cite{shime2021right}. The ICU event streams are characterised by extremely frequent measurements, high event density, and the presence of severe or anomalous events~\cite{johnson2016mimic}. 
Most articles focus on the prediction of risk, including the prediction of mortality, readmission, and the early risk of diseases such as sepsis and heart failure. For example, previous work has used the Transformer to aggregate visit-level embeddings for the prediction of heart failure at the next visit~\cite{zhang2022adadiag}, point process models to model the intensity function for mortality and septic shock~\cite{karami2023point}, and RNNs that extract sequential hidden states for in-hospital mortality and abnormal laboratory test detection~\cite{wang2019predictive}. Other works have explored CNN-RNN hybrids that combine local feature extraction with temporal aggregation to predict mortality at various time points during an ICU stay~\cite{liu2022integrated}, and attention-based models that compute weighted sums of historical events for readmission from the ICU and the onset of sepsis~\cite{shickel2022multi, barren2021improving}. 
Another major application is multi-label diagnosis and medication recommendation, where models are designed to predict the set of diagnoses or medications in the future. \citeauthor{mei2024collaborative}~\cite{mei2024collaborative} developed a Transformer model to capture dependencies between co-occurring codes for next-visit prediction, while others~\cite{lu2022multi, gao2023prescription} have addressed multi-label diagnosis and prescription recommendation using RNN-based and hybrid models that map the final hidden state to a multi-label output space. 

As clinical conditions fluctuate rapidly and interventions have immediate consequences in ICUs, the application of generating plausible future event sequences is critical for advanced modelling. This capability is used primarily for counterfactual simulation, as demonstrated by \citeauthor{wang2023style}~\cite{wang2023style}. By modelling "what-if" scenarios, these systems allow clinicians to estimate how a patient's trajectory might diverge under varying intervention strategies, providing a computational basis for personalised treatment planning. Furthermore, trajectory generation addresses the stringent privacy constraints inherent in ICU data. By creating synthetic patient records that preserve the statistical fidelity of real populations without exposing sensitive information, these methods facilitate the safe development and benchmarking of clinical algorithms across institutions~\cite{lu2023multi, theodorou2025improving}.

\textbf{Emergency Department\quad} The emergency department (ED) is the main entry point for patients who need urgent or unscheduled care~\cite{van2024machine}. EDs are characterised by short event streams, a large volume of multimodal data, and significant variability in patient presentations and acuity levels~\cite{shin2025triage}. 
In the emergency department, where rapid and precise diagnosis is paramount, the primary focus of research is the prediction of acute clinical events and early diagnosis. The authors in~\cite{de2022attentionhcare, heyman2024novel} leverage event sequences to identify imminent risks, such as opioid poisoning, sepsis, acute heart failure, and pneumonia, thereby assisting clinicians in prioritising patients for immediate intervention. 
Beyond clinical decision support, significant attention is directed towards optimising operational efficiency and patient flow. Given the frequent overcrowding in the emergency department, predicting workflow outcomes becomes critical. The authors in~\cite{hansen2023patient, van2025putting} address this by forecasting patient length-of-stay (LoS) and detecting deviations in clinical workflows. These applications provide vital intelligence for resource allocation, bed management, and the maintenance of standard operating protocols within the fast-paced ED environment.

\textbf{Hospital/Secondary Care\quad} The hospital and secondary care scenario involves the management and treatment of hospitalised patients, which includes a wide spectrum of conditions, procedures, and interventions~\cite{rough2020predicting}. In-hospital event streams are characterised by hourly records and sequences that span a few days or weeks. 
A major application of event stream modelling in hospitals is the prediction of mortality, readmission, and disease risk. Previous work has predicted in-hospital mortality~\cite{lauritsen2020early}, 30-day readmission~\cite{min2019predictive}, acute kidney injury~\cite{tomavsev2019clinically}, and the onset of chronic diseases~\cite{xiang2019time}. These models employ recurrent or attention-based encoders to transform the entire patient history into a single risk vector to enable early identification of high-risk patients and support timely clinical interventions. 
Diagnosis and medication recommendation represent another key area, where event stream models are used to predict future diagnoses, comorbidities, or medication orders for future visits~\cite{bai2018interpretable}. Previous work has applied RNN-based~\cite{suo2018multi} and attention-based~\cite{zhang2022forecasting} models to capture longitudinal dependencies by projecting the aggregated context into a multi-label output space, facilitating more accurate and personalised care planning. Generative modelling and next-event forecasting have also been explored. 
Clustering of patients has been studied using graph-based methods~\cite{jiang2020learning, niu2024ehr} and embedding-based methods~\cite{nagamine2022data}. These approaches identify clinically similar patients based on their embeddings of the event stream, supporting cohort identification, risk stratification, and tailored treatment recommendations.

\textbf{Primary Care\quad} Primary care serves as the initial point of contact and the central hub for patient management, covering a wide spectrum of preventive, acute, and chronic health services~\cite{shi2012impact}. In this context, data are usually recorded at the day-level granularity, span years, and exhibit high variability in event frequency due to the irregular nature of patient visits~\cite{blanco2019incivility}. 
Existing work has focused primarily on the prediction of chronic disease onset and the alignment of multimodal information. For example, the authors in~\cite{li2022hi, choi2017using} use Transformer and RNN to aggregate longitudinal patterns from historical event sequences and to predict whether a patient will develop a target disease, such as heart failure, within a specified time frame following a baseline date. In addition, \citeauthor{seinen2025using}~\cite{seinen2025using} has used embedding-based approaches that map diverse data types into a shared semantic space to quantify the complementarity of information across different EHR modalities, including structured records and clinical notes. These studies demonstrate the feasibility of applying event stream models to large-scale primary care datasets, supporting both risk prediction and the integration of heterogeneous data sources to improve continuity and quality of care.

\textbf{Summary\quad} The scenarios most widely studied are critical care and hospital/secondary care, reflecting the high availability of rich EHR datasets such as MIMIC~\cite{johnson2023mimic}. The ED scenario also has publicly available datasets, although the multimodal and variable nature of ED data presents unique modelling challenges~\cite{shin2025triage}. Primary care remains less explored, likely due to the scarcity of large-scale, high-quality datasets that capture extended patient histories across diverse settings. 
Regarding clinical applications, the existing literature is predominantly anchored in prediction tasks, most notably mortality, readmission, and disease onset~\cite{rajkomar2018scalable, xu2023transehr}. Although these objectives are clinically vital, they often reduce complex patient trajectories to single static outcomes, thereby underutilising the intrinsic capabilities of event stream modelling. The distinct advantage of this paradigm lies in its ability to capture the continuous temporal evolution and granular dependencies between specific clinical events—nuances that are typically lost in aggregate prediction. Consequently, there is a significant but under-explored opportunity to move beyond binary or multi-class risk stratification. Recently, some work has begun to bridge this gap by addressing dynamic, process-oriented tasks, including estimating causal treatment effect~\cite{kalia2024causal}, optimising clinical pathway~\cite{funkner2017data}, and precise scenario-specific forecasting~\cite{zheng2025scoping}.

\section{Open Research Challenges}
\label{sec:challenges}

\textbf{Co-occurring Events\quad} A fundamental challenge in modelling EHR event streams is the handling of co-occurring clinical events. Although sequential architectures such as RNNs and Transformers presuppose strict temporal order ($t_1 < t_2 < \dots < t_n$), clinical workflows frequently generate multiple observations recorded at the same timestamp~\cite{shou2023concurrent}. These simultaneous events constitute an unordered set, and their relative arrangement in sequences is clinically irrelevant. This creates a structural conflict with standard deep learning models: to process these events, they must be forced into an arbitrary sequence, which risks the model incorrectly interpreting this artificial order as a meaningful temporal dependency. 
Existing approaches rely on two main strategies to accommodate this modality, each with notable drawbacks. The first strategy aggregates co-occurring events into a single timestep representation, using techniques like multi-hot encoding or composite tokens~\cite{cheng2016risk, hur2020facilitating, coombes2022sequences, henriksson2016ensembles}. However, this aggregation inherently leads to information loss, as it conflates clinical entities into a single vector and often discards the specific values associated with individual event types. 
The second strategy linearises the set by imposing an artificial order, often employing random shuffling or permutation-invariant layers to mitigate order sensitivity~\cite{zhang2020diagnostic, liu2020learning, rough2020predicting, li2022next, ma2024memorize}. Despite these efforts, imposing an artificial sequence can lead to divergent representations and spurious causal inferences. For example, event stream models can incorrectly take a systolic blood pressure as the cause of a simultaneous diastolic reading due to its position~\cite{jurewicz2021set}. Moreover, this transformation exponentially expands the input space (a set of $m$ events admits $m!$ permutations), which complicates the learning task especially for complex clinical patterns.

\textbf{Long Context\quad} EHRs can span decades, resulting in event streams that far exceed the context window limits of standard deep learning architectures~\cite{wornow2025context, zhu2025medtpe}. While many traditional approaches aggregate history into coarser representations, event stream modelling aims to preserve the granularity of individual events as tokens in a timeline. This leads to sequences that can easily reach hundreds of thousands of tokens, making standard attention mechanisms computationally intractable at scale~\cite{zhu2025medtpe}. 
This limitation forces a trade-off between history and granularity. Simply truncating the sequence to fit the model's input limit risks discarding long-term historical context, which is crucial to understanding the patient's baseline risk~\cite{wornow2025context}. In contrast, prioritising long-term history may require compressing or down-sampling recent acute events, potentially obscuring immediate clinical deterioration~\cite{zhu2025medtpe}. Developing mechanisms to efficiently balance the representation of distant history with recent high-frequency events remains an open challenge.

\textbf{The Cold-Start Problem and Information Sufficiency\quad} A critical limitation of event stream models is their reliance on extensive historical context to generate robust representations. In real-world clinical practice, healthcare systems frequently encounter the "cold-start" problem: patients presenting for the first time with little to no prior recorded medical history~\cite{tan2022metacare++}. While architectures such as Transformers excel at finding patterns across long patient journeys, their performance often degrades significantly when forced to make high-stakes predictions based on highly sparse, single-visit data~\cite{li2022hi, tan2022metacare++}.
This raises the unresolved question of information sufficiency: how much data is sufficient for a model to make reliable clinical inferences? Existing literature often side-steps this issue by applying arbitrary minimum-length thresholds for patient inclusion during dataset preprocessing, effectively excluding the population with limited data from the training and evaluation phases~\cite{van2022developing}. Consequently, models may become over-reliant on the sheer volume of historical events rather than the acute, immediate signals critical for early intervention. Developing architectures that can dynamically quantify their own uncertainty and maintain robust predictive performance even at the very beginning of a patient's clinical trajectory remains a major open challenge~\cite{chen2025uncertainty}.

\textbf{Error Accumulation and Clinical Plausibility\quad} A key advantage of event stream modelling is its ability to generate potential future patient trajectories. However, while this generative capability is highly valuable for supporting clinical decision-making, it introduces the dual challenges of error accumulation~\cite{bengio2015scheduled} and clinical plausibility~\cite{biswal2021eva}. Because these autoregressive models predict future states sequentially, an erroneous prediction generated in the middle of the event stream becomes the conditioning context for all subsequent steps. This phenomenon causes errors to cascade exponentially, leading to generated trajectories that rapidly diverge from realistic patient outcomes over longer time horizons~\cite{bengio2015scheduled}.
Furthermore, deep learning models operate primarily as statistical pattern matchers and lack an inherent understanding of physiological rules, biological limits, or strict clinical workflows~\cite{song2022incorporating, sirocchi2024medical}. Without explicit constraints, a model might generate a sequence that violates physical reality or medical logic. For example, the model may generate a negative blood pressure reading, predict a hospital discharge event prior to an admission, or suggest a pregnancy test for a male patient. While some approaches attempt to mitigate this through post-hoc filtering, integrating hard physiological constraints and deterministic clinical rules directly into the generative modelling process is essential to ensure that predicted trajectories remain physically viable and clinically safe~\cite{sirocchi2024medical}.

\textbf{The Absence of Standardised Benchmarking\quad} Despite the rapid proliferation of EHR event stream architectures, the field is currently hindered by a distinct lack of standardised benchmarking protocols~\cite{chen2026cross}. As evidenced by the evaluation strategies summarised in Appendix Table~\ref{append_table: Summary of papers}, the existing literature employs a highly fragmented array of performance metrics. For identical downstream tasks, such as risk prediction or multi-label diagnosis, studies frequently diverge in their choice of evaluation metrics, alternating between the area under the receiver operating characteristic curve (AUC), area under the precision-recall curve (AUPRC), F1-score, and Mean Average Precision (MAP). Furthermore, while public datasets like MIMIC-III~\cite{johnson2016mimic} and MIMIC-IV~\cite{johnson2023mimic} are commonly utilised, a significant proportion of research relies on private, institutional cohorts. This dependence on private data, compounded by inconsistent metric selection, creates a disjointed evaluation landscape that precludes direct, equitable comparisons between models. Consequently, it is often impossible to determine whether a model's reported superiority stems from genuine algorithmic innovation, specific data preprocessing choices, or favourable metric selection~\cite{beam2020challenges}. To mature, the field urgently requires a unified, open-source benchmark comprising standardised clinical datasets, consistent event stream curation pipelines, and fixed evaluation metrics tailored to specific clinical scenarios to rigorously assess the true efficacy of these models.

\section{Future Directions}
\label{sec:future_directions}

\subsection{Multimodal Event Streams}
Although current research largely treats structured data and unstructured notes as separate inputs fused at a late stage, the event stream representation offers a unique opportunity to unify these modalities into a single, chronologically consistent timeline. Future work should move beyond simple concatenation and instead treat complex data types, including clinical narratives, images, and bio-signals, as distinct clinical events $(t, c, v)$ within the stream. The primary challenge and opportunity lie in feature alignment: mapping high-frequency continuous signals (e.g., biosensors) and sparse, information-dense events (e.g., MRI scans) into the same discrete event sequence $X = \{X_q\}$. 
Explicitly modelling these modalities as interleaved events allows the model to capture the causal interplay between different data types, for example, how a sudden drop in a biosensor reading ($t_i$) triggers a specific imaging order ($t_{i+1}$) and a subsequent diagnosis code ($t_{i+2}$)~\cite{mcdermott2023event}. Approaches that use unified set embeddings or bottleneck tokens to exchange information between modality-specific encoders at each timestep are particularly promising~\cite{lee2023learning}. Furthermore, methods that align embeddings from disparate sources can resolve the ambiguity of sparse structured records, providing a denser, more context-rich trajectory for downstream tasks~\cite{wang2024multimodal, lu2022multi, he2020attention, ding2024distilling, lin2022mr, seinen2025using, li2022next}.

\subsection{Large Event Models}
\label{sec:foundation_models}
The sequential nature of EHR event streams makes them the ideal substrate for developing large event stream models (LEMs), the foundation models pre-trained specifically for clinical events~\cite{guo2024multi}. Unlike tabular models, event stream architectures can leverage next-event prediction objectives (predicting the time $t_{n+1}$, type $c_{n+1}$, and value $v_{n+1}$) to learn the intrinsic temporal dynamics of disease progression. 
The future evolution in this domain is likely to follow two primary trajectories. The first is semantic integration, where models leverage LLMs to ground abstract clinical codes in natural language. By translating structured event tuples $(t, c, v)$ into textual descriptions, these systems can process the patient's history as a coherent narrative~\cite{mcdermott2023event}. This capability enables the interpretation of rare codes through their definitions and facilitates the generation of transparent, natural language explanations for model predictions~\cite{wu2024instruction, agarwal2024faithfulness, ma2024memorize}. 
The second frontier is generative trajectory modelling, which shifts the application from risk prediction to trajectory simulation. This direction focuses on learning the complex probability distributions of event sequences to construct a "digital twin" of the patient, which is plausible future timelines given the current patient's health status~\cite{kuruppu2025health, wang2024twin}. By simulating how a patient's path evolves, these generative models enable robust counterfactual analysis. For instance, clinicians can simulate "what-if" scenarios to observe how a trajectory might diverge under different treatment protocols. Furthermore, this generative capacity facilitates the creation of high-fidelity synthetic data, supporting research and decision-making while rigorously preserving patient privacy~\cite{yoon2023ehr, wang2023style}.

\subsection{Reinforcement Learning for Optimisation}
Although the current landscape of EHR event stream modelling is predominantly focused on predictive tasks and trajectory generation~\cite{amirahmadi2023deep}, a critical future frontier lies in optimisation via reinforcement learning (RL). The chronological sequence of clinical events and interventions of event streams maps naturally onto the Markov decision process~\cite{yu2021reinforcement}. This natural alignment renders the event stream representation an ideal substrate for offline RL algorithms to optimise the prediction and generation capabilities of the event stream models. 
One promising direction is utilising the robust, time-aware embeddings generated by self-supervised event models as the continuous state space for RL agents~\cite{steinberg2024motor}. Rather than merely forecasting future events, these agents could recommend optimal sequences of interventions to maximise long-term patient health outcomes while minimising adverse events. In addition, RL techniques can be integrated into the generative modelling process itself to optimise the overall quality and clinical plausibility of the predicted patient trajectories~\cite{wang2024esrl}. By formulating appropriate reward functions, models can be explicitly penalised for generating physically impossible events or violating established medical workflows, while receiving high rewards for producing coherent, biologically viable sequences. 
\section{Conclusion}
\label{sec:conclusion}

This review establishes a unified framework for EHR event stream modelling by categorising methods based on their handling of time, type, and value. A significant trend identified is the methodological shift from task-specific supervised training to self-supervised learning that leverages longitudinal unlabelled data to learn robust representations. Applications across critical care, ED, hospital/secondary care, and primary care demonstrate the versatility of event stream models in supporting tasks such as diagnosis prediction, risk prediction, and trajectory generation. However, the field still faces significant challenges, particularly in handling co-occurring events, managing long temporal contexts, dealing with the cold-start problem, addressing the challenges of error accumulation and clinical plausibility, and the absence of standardised benchmarking. Looking forward, future research can leverage the event stream as a unified multimodal representation to integrate structured and unstructured data, develop the generative large-event models for patient trajectory simulation, and optimise the event stream models with RL. This evolution positions the EHR event stream as a foundation for the next generation of personalised clinical decision support.

\bibliographystyle{apalike}
\bibliography{reference}

\clearpage

\appendix

\section{Summary of EHR Event Stream Modelling}

\begin{center}
    \tablecaption{Summary of EHR event stream modelling methods with supervised learning.}
    \label{append_table: Summary of papers}
    % \tablefirsthead{
    % }
    
    \tablehead{
      \hline
        \multirow{2}{*}{Reference} 
        & \multicolumn{2}{c}{\textbf{Event Stream Modelling}}
        & \multicolumn{2}{c}{\textbf{Training Strategy}}
        & \multicolumn{4}{c}{\textbf{Applications}} \\ 
        \cline{2-3} \cline{4-5} \cline{6-9} 
        & Time & Type \& Value 
        & Supervised & Self-Sup. 
        & Scenario & Task & Metric & Dataset \\ 
      \hline
    }
    
    \tabletail{
        \hline
        \multicolumn{3}{r}{continued on next page} \\
        \\
    }
    
    \tablelasttail{
        \hline
        \multicolumn{3}{r}{end of the table} \\
        \\
    }
    
    \renewcommand{\arraystretch}{1.1}
    \tiny
    \begin{supertabular}{c cc cc p{1.2cm} p{2cm}p{1.8cm}p{2cm}}
    \hline
        \multirow{2}{*}{Reference} 
        & \multicolumn{2}{c}{\textbf{Event Stream Modelling}}
        & \multicolumn{2}{c}{\textbf{Training Strategy}}
        & \multicolumn{4}{c}{\textbf{Applications}} \\ 
        \cline{2-3} \cline{4-5} \cline{6-9} 
        & Time & Type \& Value 
        & Supervised & Self-Sup. 
        & Scenario & Task & Metric & Dataset \\ 
      \hline
%%%%%%%%%%%%%%%%%%%%%%%% 2016-2017 %%%%%%%%%%%%%%%%%%%%%%%%%%%%%%%
\multicolumn{2}{c}{\textbf{2016 Year:}} \\
\hline
\cite{cheng2016risk} & Feature & Type Only & $\checkmark$ & - & Hospital & Risk Prediction & AUC & IBM Warehouse \\
\hline
\cite{henriksson2016ensembles} & Feature & Type Only & - & Skip-gram & Hospital & Risk Prediction & ACC, AUC & Stockholm EPR \\
\hline

\multicolumn{2}{c}{\textbf{2017 Year:}} \\
\hline
\cite{choi2017using} & Feature & Type Only & $\checkmark$ & - & Primary Care & Diagnosis & AUC & Sutter Palo Alto \\
\hline
\cite{weiss2017piecewise} & Feature & Type Only & $\checkmark$ & - & ICU & Risk Prediction & LL & MIMIC-III \\
\hline
%%%%%%%%%%%%%%%%%%%%%%%% 2018 %%%%%%%%%%%%%%%%%%%%%%%%%%%%%%%
\multicolumn{2}{c}{\textbf{2018 Year:}} \\
\hline

      \cite{wu2018modeling} & Feature & Type Only & $\checkmark$ & - & ICU & Risk Prediction & PRE, REC, F1, AUC & PhysioNet-2012 \\
\hline
\multirow{2}{*}{\cite{bai2018interpretable}} & \multirow{2}{*}{Feature} & \multirow{2}{*}{Type Only} & \multirow{2}{*}{$\checkmark$} & \multirow{2}{*}{-} & \multirow{2}{*}{Hospital} & \multirow{2}{*}{Diagnosis} & \multirow{2}{*}{ACC, W-F1} & Private \\
 & & & & & & & & Mixed Dataset \\
 \hline
 \cite{liu2018learning} & Feature & Type Only & $\checkmark$ & - & ICU & Risk Prediction & AUC, AP & MIMIC-III \\
 \hline
 \multirow{2}{*}{\cite{suo2018multi}} & \multirow{2}{*}{Feature} & \multirow{2}{*}{Type Only} & \multirow{2}{*}{$\checkmark$} & \multirow{2}{*}{-} & \multirow{2}{*}{Hospital} & \multirow{2}{*}{Risk Prediction} & \multirow{2}{*}{ACC} & SOF \\
 & & & & & & & & BloodTest \\
\hline
\cite{xu2018learning} & Feature & Type Only & $\checkmark$ & - & Hospital & \shortstack{Patient Clustering, \\ LoS} & NMI, R@N, AUPRC & New Rural Co-op \\
\hline
  \cite{qiao2018pairwise} & Feature & Type \& Value & $\checkmark$ & - & ICU & \shortstack{Risk Prediction} & R@K, RMSE & MIMIC-III \\
\hline

%%%%%%%%%%%%%%%%%%%%%%%% 2019 %%%%%%%%%%%%%%%%%%%%%%%%%%%%%%%
\multicolumn{2}{c}{\textbf{2019 Year:}} \\
\hline
  \cite{yin2019domain} & Feature & Type Only & $\checkmark$ & - & ICU & Risk Prediction & AUC & MIMIC-III \\
\hline
\cite{wang2019predictive} & Feature & Type Only & $\checkmark$ & - & ICU & Risk Prediction & AUC, AP & MIMIC-III \\
\hline
\cite{fu2019pearl} & Feature & Type Only & $\checkmark$ & - & ICU & Risk Prediction & AUC & MIMIC-III \\
\hline
\cite{yao2019multiple} & Feature & Type Only & $\checkmark$ & - & Primary Care, ED, Hospital & Risk Prediction & AUC, ACC, F1 & NHIRD \\
\hline
\cite{zhang2019interpretable} & Feature & Type Only & $\checkmark$ & - & ICU & Risk Prediction & AUC & MIMIC-III \\
\hline
\cite{min2019predictive} & Feature & Type Only & $\checkmark$ & - & Hospital & Risk Prediction & AUC & Geisinger \\
\hline
\cite{tomavsev2019clinically} & Feature & Type Only & $\checkmark$ & - & Hospital & Risk Prediction & AUC, AUPRC & VA EHR \\
\hline
\multirow{2}{*}{\cite{huang2019time}} & \multirow{2}{*}{Feature} & \multirow{2}{*}{Type Only} & \multirow{2}{*}{$\checkmark$} & \multirow{2}{*}{-} & ICU & \multirow{2}{*}{Risk Prediction} & \multirow{2}{*}{\shortstack{ACC, AUPRC, \\ AUC}} & MIMIC-III \\
 & & & & & Hospital & & & PPMI \\
\hline

\cite{ruan2019representation} & Feature & Type Only & - & Next-tok & Hospital & Risk Prediction & AUC & Shuguang Hosp. \\
\hline
\cite{xiang2019time} & Feature & Type Only & - & Skip-gram & ED, Hospital, ICU & Risk Prediction & AUC & Cerner Health Facts \\
\hline

%%%%%%%%%%%%%%%%%%%%%%%% 2020 %%%%%%%%%%%%%%%%%%%%%%%%%%%%%%%
\multicolumn{2}{c}{\textbf{2020 Year:}} \\
\hline
\cite{li2020marrying} & Feature & Type Only & $\checkmark$ & - & Combined & Risk Prediction & AUC, SEN, SPE & KnowLife \\
\hline
\cite{lauritsen2020early} & Feature & Type Only & $\checkmark$ & - & Hospital & Risk Prediction & AUC, mAP, NB & Private \\
\cite{lu2020clinical} & Feature & Type Only & $\checkmark$ & - & ICU & Diagnosis & \shortstack{PRE, R@k, \\ MAP@k, AUC} & MIMIC-III \\
\hline
\multirow{3}{*}{\cite{he2020attention}} & \multirow{3}{*}{Feature} & \multirow{3}{*}{Multimodal} & \multirow{3}{*}{$\checkmark$} & \multirow{3}{*}{-} & ICU & \multirow{3}{*}{\shortstack{Medication Rec, DRG}} & \multirow{3}{*}{\shortstack{JSC, M-F1, \\ AUPRC, ACC}} & MIMIC-III \\
 & & & & & Hospital & & & Chinese Hospital \\
 & & & & & Non-clinical & & & Tax Agency \\
\hline
\cite{liu2020learning} & Feature & Type Only & $\checkmark$ & - & ICU & Risk Prediction & AUC, AUPRC & MIMIC-III \\
\hline
\cite{hur2020facilitating} & Feature & Type Only & $\checkmark$ & - & ICU & \shortstack{Diagnosis , \\ Risk Prediction} & AUC, F1 & MIMIC-III \\
\hline
\cite{wang2020ehr2vec} & Feature & Type Only & $\checkmark$ & - & Hospital & Medication Rec. & ACC & SLE Clinical Notes \\
\hline
\cite{rough2020predicting} & Feature & Type Only & $\checkmark$ & - & Hospital & Medication Rec. & R@K, AUC, AUPRC & UCSF EHR \\
\hline
\cite{peng2020self} & Feature & Type Only & $\checkmark$ & - & ICU & \shortstack{Diagnosis, Risk Prediction} & AUPRC, P@k & MIMIC-III \\
\hline
\cite{hettige2020medgraph} & Feature & Type Only & $\checkmark$ & - & Hospital & Risk Prediction & AUC, AP, F1 & Proprietary \\
\hline
\cite{xiang2020asthma} & Feature & Type Only & $\checkmark$ & - & ED, Hospital, ICU & Risk Prediction & AUC & Cerner Health Facts \\
\hline
\cite{cho2020process} & Feature & Type Only & $\checkmark$ & - & ED & Triage & \shortstack{ACC, PRE, \\ REC, F1} & Korean Tertiary ER \\
\hline
\cite{ljubic2020predicting} & Feature & Type Only & $\checkmark$ & - & Hospital & Risk Prediction & ACC, SEN, SPE & HCUP SID \\
\hline
\multirow{2}{*}{\cite{jiang2020learning}} & \multirow{2}{*}{Feature} & \multirow{2}{*}{Type Only} & \multirow{2}{*}{$\checkmark$} & \multirow{2}{*}{-} & Hospital & Patient Retrieval & nDCG, HL & Private Dataset \\
 & & & & & ICU & Patient Clustering & SSE, NMI, AUC & MIMIC-III \\
\hline
\cite{lee2020multi} & Feature & Type Only & - & Next-tok & ICU & Trajectory Generation & AUPRC & MIMIC-III \\
\hline
\cite{lin2020medical} & Feature & Type Only & - & Skip-gram & ICU & \shortstack{Patient Clustering} & \shortstack{HRR, AUC, \\ ACC, NMI} & MIMIC-III \\
\hline
\cite{zhang2020diagnostic} & Feature & Type Only & - & Contrastive & ICU & Diagnosis & AUC, AUPRC & MIMIC-III \\
\hline

%%%%%%%%%%%%%%%%%%%%%%%% 2021 %%%%%%%%%%%%%%%%%%%%%%%%%%%%%%%
\multicolumn{2}{c}{\textbf{2021 Year:}} \\
\hline
\cite{men2021multi} & Feature & Type Only & $\checkmark$ & - & Hospital & Diagnosis & PRE, REC, F1, WAUC & Southeast China EHR \\
\hline
\cite{tomavsev2021use} & Feature & Type Only & $\checkmark$ & - & Hospital & Risk Prediction & AUC, AUPRC & VA EHR \\
\hline
\cite{dong2021identifying} & Feature & Type Only & $\checkmark$ & - & ED, Hospital, ICU & Diagnosis & \shortstack{F1, PRE, \\ REC, AUC} & Cerner Health Facts \\
\hline
\cite{kwak2021interpretable} & Feature & Type Only & $\checkmark$ & - & Hospital & Risk Prediction & AUC, AUPRC & NHIS-NSC \\
\hline
\cite{barren2021improving} & Feature & Type Only & $\checkmark$ & - & ICU & Risk Prediction & AUPRC, AUC & MIMIC-III \\
\hline
\cite{rethmeier2021efficare} & Feature & Type Only & $\checkmark$ & - & ICU & \shortstack{Risk Prediction \\, Operational} & \shortstack{AUC, AUPRC, \\ Kappa} & MIMIC-III \\
\hline
\cite{gao2021causal} & Feature & Type Only & $\checkmark$ & - & Hospital & Causal Inference & RMSE & UCI Diabetes \\
\hline
\cite{zhang2021synteg} & Feature & Type Only & - & Next-tok & Hospital & Trajectory Generation & AUC & Vanderbilt SD \\
\hline
\cite{pang2021cehr} & Feature & Type Only & - & Masked & Hospital & Risk Prediction & AUC, AUPRC & CUIMC-NYP \\
\hline

%%%%%%%%%%%%%%%%%%%%%%%% 2022 %%%%%%%%%%%%%%%%%%%%%%%%%%%%%%%
\multicolumn{2}{c}{\textbf{2022 Year:}} \\
\hline
\multirow{2}{*}{\cite{ru2022sparse}} & \multirow{2}{*}{Feature} & \multirow{2}{*}{Type \& Value} & \multirow{2}{*}{$\checkmark$} & \multirow{2}{*}{-} & \multirow{2}{*}{ICU} & \multirow{2}{*}{\shortstack{Trajectory Generation, \\Diagnosis}} & \multirow{2}{*}{\shortstack{F1, RMSE, \\ M-PRE/REC}} & MIMIC-II \\
 & & & & & & & & MIMIC-III \\
\hline
\cite{liu2022modeling} & Feature & Type Only & $\checkmark$ & - & ICU & Risk Prediction & AUC, AUPRC & MIMIC-III \\
\hline
\multirow{2}{*}{\cite{liu2022integrated}} & \multirow{2}{*}{Feature} & \multirow{2}{*}{Type Only} & \multirow{2}{*}{$\checkmark$} & \multirow{2}{*}{-} & \multirow{2}{*}{ICU} & \multirow{2}{*}{Risk Prediction} & \multirow{2}{*}{AUC, AUPRC} & eICU \\
 & & & & & & & & MIMIC-III \\
\hline
\cite{sengupta2022analyzing} & Feature & Type Only & $\checkmark$ & - & Hospital & Risk Prediction & AUC & N3C \\
\hline
\multirow{2}{*}{\cite{de2022attentionhcare}} & \multirow{2}{*}{Feature} & \multirow{2}{*}{Type Only} & \multirow{2}{*}{$\checkmark$} & \multirow{2}{*}{-} & ICU & \multirow{2}{*}{Diagnosis} & \multirow{2}{*}{R@k, P@k, AUC, F1} & MIMIC-III \\
 & & & & & ED & & & MIMIC-IV-ED \\
\hline
\cite{song2022tri} & Feature & Type \& Value & $\checkmark$ & - & ICU & Trajectory Generation & LL, ACC, RMSE & MIMIC-II \\
\hline
\cite{zhu2022temporal} & Feature & Type Only & $\checkmark$ & - & ICU & Diagnosis & MAP, F1, P@1 & MIMIC-III \\
\hline
\cite{lu2022multi} & Feature & Multimodal & $\checkmark$ & - & ICU & Diagnosis & PRE, REC, F1 & MIMIC-III \\
\hline
\cite{shickel2022multi} & Feature & Type Only & $\checkmark$ & - & ICU & Risk Prediction & AUC & UFHealth \\
\hline
\cite{coombes2022sequences} & Feature & Type \& Value & $\checkmark$ & - & ICU & Medication Rec, Risk Prediction & AUC & MIMIC-III \\
\hline
\cite{ghanzouri2022performance} & Feature & Type Only & $\checkmark$ & - & Hospital, ICU & Risk Prediction & AUC & STARR \\
\hline
\multirow{2}{*}{\cite{hong2022cd}} & \multirow{2}{*}{Feature} & \multirow{2}{*}{Type Only} & \multirow{2}{*}{$\checkmark$} & \multirow{2}{*}{-} & \multirow{2}{*}{ICU} & \multirow{2}{*}{Risk Prediction} & \multirow{2}{*}{C-idx, AUC, ACC} & MIMIC-III \\
 & & & & & & & & PLAGH HF \\
\hline
\cite{li2022next} & Feature & Multimodal & $\checkmark$ & - & Hospital & Operational & R@K, NDCG@K & Chinese Tertiary \\
\hline
\multirow{2}{*}{\cite{zhang2022adadiag}} & \multirow{2}{*}{Feature} & \multirow{2}{*}{Type Only} & \multirow{2}{*}{-} & \multirow{2}{*}{Masked} & ICU & \multirow{2}{*}{Diagnosis} & \multirow{2}{*}{AUC, AUPRC} & MIMIC-IV \\
 & & & & & Hospital & & & UCLA \\
\hline
\cite{lin2022mr} & Feature & Multimodal & - & Masked & Hospital & Medication Rec. & JSC, F1, AUPRC & Private \\
\hline
\cite{li2022hi} & Feature & Type Only & - & Contrastive & Primary Care & Risk Prediction & AUC, AUPRC & CPRD \\
\hline
\multirow{2}{*}{\cite{zhang2022forecasting}} & \multirow{2}{*}{Feature} & \multirow{2}{*}{Type Only} & \multirow{2}{*}{-} & \multirow{2}{*}{Contrastive} & \multirow{2}{*}{Hospital} & \multirow{2}{*}{Diagnosis} & \multirow{2}{*}{AUC, AUPRC} & VUMC \\
 & & & & & & & & All of Us \\
\hline

%%%%%%%%%%%%%%%%%%%%%%%% 2023 %%%%%%%%%%%%%%%%%%%%%%%%%%%%%%%
\multicolumn{2}{c}{\textbf{2023 Year:}} \\
\hline
\cite{antikainen2023transformers} & Feature & Type Only & $\checkmark$ & - & Hospital & Risk Prediction & AUC, PRE, REC, SPE & Tays Heart Hospital \\
\hline
\cite{wang2023style} & Feature & Type Only & $\checkmark$ & - & ICU & Risk Prediction & ACC, VAL & MIMIC-III \\
\hline
\cite{shou2023concurrent} & Feature & Type Only & $\checkmark$ & - & ICU & Diagnosis & W-F1, WAUC & MIMIC-III \\
\hline
\multirow{2}{*}{\cite{yang2024dynamic}} & \multirow{2}{*}{Feature} & \multirow{2}{*}{Type Only} & \multirow{2}{*}{$\checkmark$} & \multirow{2}{*}{-} & \multirow{2}{*}{ICU} & \multirow{2}{*}{Diagnosis} & \multirow{2}{*}{ACC} & MIMIC-III \\
 & & & & & & & & MIMIC-IV \\
\hline
\cite{gao2023prescription} & Feature & Type Only & $\checkmark$ & - & ICU & Medication Rec. & F1, AUPRC, AP & MIMIC-III \\
\hline
\cite{hansen2023patient} & Feature & Type Only & $\checkmark$ & - & ED & Operational (LoS) & AUC, F1, MSE & Danish EHR \\
\hline
\multirow{3}{*}{\cite{lee2023learning}} & \multirow{3}{*}{Feature} & \multirow{3}{*}{Multimodal} & \multirow{3}{*}{$\checkmark$} & \multirow{3}{*}{-} & \multirow{3}{*}{ICU} & \multirow{3}{*}{Risk Prediction} & \multirow{3}{*}{AUC, AUPRC} & MIMIC-IV \\
 & & & & & & & & MIMIC-CXR \\
 & & & & & & & & MIMIC-Note \\
\hline
\cite{li2023prediction} & Feature & Type Only & $\checkmark$ & - & ED, Hospital, ICU & Risk Prediction & AUC & Cerner Health Facts \\
\hline
\multirow{2}{*}{\cite{wu2023iterative}} & \multirow{2}{*}{Feature} & \multirow{2}{*}{Type \& Value} & \multirow{2}{*}{$\checkmark$} & \multirow{2}{*}{-} & \multirow{2}{*}{ICU} & \multirow{2}{*}{Risk Prediction} & \multirow{2}{*}{AUPRC, AUC} & eICU \\
 & & & & & & & & MIMIC-IV \\
\hline
\multirow{2}{*}{\cite{karami2023point}} & \multirow{2}{*}{Feature} & \multirow{2}{*}{Type Only} & \multirow{2}{*}{-} & \multirow{2}{*}{Masked} & \multirow{2}{*}{ICU} & \multirow{2}{*}{\shortstack{Risk Prediction}} & \multirow{2}{*}{AUC, AUPRC} & PhysioNet-2012 \\
 & & & & & & & & PhysioNet-2019 \\
\hline
\multirow{3}{*}{\cite{xu2023transehr}} & \multirow{3}{*}{Feature} & \multirow{3}{*}{Type \& Value} & \multirow{3}{*}{-} & \multirow{3}{*}{Masked} & \multirow{3}{*}{ICU} & \multirow{3}{*}{\shortstack{Risk Prediction, \\ Phenotyping}} & \multirow{3}{*}{\shortstack{AUC, AUPRC, \\ MAE, $\kappa$}} & PhysioNet-2012 \\
 & & & & & & & & MIMIC-III \\
 & & & & & & & & AmsterdamUMCdb \\
\hline
\multirow{2}{*}{\cite{lu2023multi}} & \multirow{2}{*}{Feature} & \multirow{2}{*}{Type Only} & \multirow{2}{*}{-} & \multirow{2}{*}{Next-tok} & ICU & \multirow{2}{*}{Trajectory Generation} & \multirow{2}{*}{ND, JSD} & MIMIC-III \\
 & & & & & ICU & & & MIMIC-IV \\
\hline
\cite{mcdermott2023event} & Feature & Type Only & - & Next-tok & Hospital & Risk Prediction & AUC & MIMIC-IV \\
\hline
\cite{yang2023transformehr} & Event & Type Only & - & Next-tok & Hospital & Diagnosis & AUC, AUPRC, PPV & MIMIC-IV \\
\hline
\cite{ramirez2023deep} & Feature & Type Only & - & Next-tok & Hospital & Trajectory Generation & ACC, F1, AUC & 4TU Sepsis \\
\hline

%%%%%%%%%%%%%%%%%%%%%%%% 2024 %%%%%%%%%%%%%%%%%%%%%%%%%%%%%%%
\multicolumn{2}{c}{\textbf{2024 Year:}} \\
\hline
\cite{sheng2024mining} & Feature & Type Only & $\checkmark$ & - & Hospital & Risk Prediction & PRE, REC, F1 & NHIRD \\
\hline
\multirow{2}{*}{\cite{mei2024collaborative}} & \multirow{2}{*}{Feature} & \multirow{2}{*}{Type Only} & \multirow{2}{*}{$\checkmark$} & \multirow{2}{*}{-} & ICU & \multirow{2}{*}{Diagnosis} & \multirow{2}{*}{W-F1, R@10} & MIMIC-III \\
 & & & & & Hospital & & & MIMIC-IV \\
\hline
\multirow{2}{*}{\cite{chen2024predictive}} & \multirow{2}{*}{Feature} & \multirow{2}{*}{Type Only} & \multirow{2}{*}{$\checkmark$} & \multirow{2}{*}{-} & ICU & \multirow{2}{*}{Diagnosis} & \multirow{2}{*}{P@k, ACC@k} & MIMIC-III \\
 & & & & & Hospital & & & MIMIC-IV \\
\hline
\cite{nigo2024deep} & Feature & Type Only & $\checkmark$ & - & Hospital & Risk Prediction & \shortstack{AUC, SPE, \\ SEN, PPV} & MIMIC-IV \\
\hline
\cite{chowdhury2024stratifying} & Feature & Type Only & $\checkmark$ & - & Hospital & Risk Prediction & RMSE & Mayo Clinic \\
\hline
\cite{heyman2024novel} & Feature & Type Only & $\checkmark$ & - & ED & Diagnosis & AUC & Halland EHR \\
\hline
\cite{shahidi2024exploring} & Feature & Type Only & $\checkmark$ & - & Primary Care, Hospital & Risk Prediction & AUC, SEN, PRE & Alberta Health \\
\hline
\cite{vabalas2024deep} & Feature & Type Only & $\checkmark$ & - & Primary Care & Risk Prediction & \shortstack{AUC, C-idx, \\ AUPRC} & Finnish Registry \\
\hline
\cite{sofiia2024graph} & Feature & Type Only & $\checkmark$ & - & Hospital & Risk Prediction & ACC, PRE, REC & Almazov Centre \\
\hline
\cite{kamruzzaman2024improving} & Feature & Type Only & $\checkmark$ & - & Hospital & Risk Prediction & \shortstack{AUC, AUPRC, \\ SEN, SPE} & UVA Hospital \\
\hline
\multirow{2}{*}{\cite{kovtun2024label}} & \multirow{2}{*}{Feature} & \multirow{2}{*}{Type Only} & \multirow{2}{*}{$\checkmark$} & \multirow{2}{*}{-} & ICU & \multirow{2}{*}{Diagnosis} & \multirow{2}{*}{W-F1, W-AUC, HL} & MIMIC-III \\
 & & & & & Synthetic & & & Synthea \\
\hline
\cite{li2024temporal} & Feature & Type Only & $\checkmark$ & - & Hospital & Pathway Mining & TC@10, NKQM@N & Chinese Univ. Hosp. \\
\hline
\cite{wu2024instruction} & Event & Type Only & - & Next-tok & ICU & Risk Prediction & AUC & MIMIC-Instr \\
\hline
\cite{kanchinadam2024large} & Feature & Type Only & - & Next-tok & Hospital & Trajectory Generation & \shortstack{AUC, F1, \\ BLEU, R2} & U.S. Insurer \\
\hline
\cite{renc2024zero} & Event & Type \& Value & - & Next-tok & Hospital & Risk Prediction & \shortstack{AUC, AUPRC, \\ MAE} & MIMIC-IV \\
\hline
\multirow{2}{*}{\cite{guo2024multi}} & \multirow{2}{*}{Feature} & \multirow{2}{*}{Type \& Value} & \multirow{2}{*}{-} & \multirow{2}{*}{Next-tok} & Hospital & \multirow{2}{*}{Risk Prediction} & \multirow{2}{*}{AUC} & MIMIC-IV \\
 & & & & & Hospital & & & SickKids \\
\hline

\cite{xie2024hgthp} & Feature & Type \& Value & - & Next-tok & ICU & Trajectory Generation & RMSE, ER & MIMIC-II \\
\hline
\multirow{2}{*}{\cite{wang2024twin}} & \multirow{2}{*}{Feature} & \multirow{2}{*}{Multimodal} & \multirow{2}{*}{-} & \multirow{2}{*}{Next-tok} & \multirow{2}{*}{Hospital} & \multirow{2}{*}{\shortstack{Trajectory Generation,\\ Risk Prediction}} & \multirow{2}{*}{\shortstack{r, AUC, \\ SEN, NNAA}} & Phase III Trial \\
 & & & & & & & & TOP \\
\hline
\cite{ding2024distilling} & Feature & Multimodal & - & Next-tok & ICU & Risk Prediction & F1, AUC, AUPRC & MIMIC-III \\
\hline
\multirow{2}{*}{\cite{karami2024tee4ehr}} & \multirow{2}{*}{Feature} & \multirow{2}{*}{Type Only} & \multirow{2}{*}{-} & \multirow{2}{*}{Masked} & \multirow{2}{*}{ICU} & \multirow{2}{*}{Risk Prediction} & \multirow{2}{*}{\shortstack{LL, F1, \\ AUC, AUPRC}} & PhysioNet-2012 \\
 & & & & & & & & PhysioNet-2019 \\
\hline
\cite{niu2024ehr} & Feature & Type Only & - & Masked & Hospital & Anomaly Detection & PRE, REC, F1 & Consults/Labs \\
\hline
\cite{oufattole2024event} & Feature & Type \& Value & - & Contrastive & Hospital & Risk Prediction & AUC, AP & MIMIC-IV \\
\hline
\multirow{2}{*}{\cite{ma2024memorize}} & \multirow{2}{*}{Feature} & \multirow{2}{*}{Type Only} & \multirow{2}{*}{-} & \multirow{2}{*}{Contrastive} & ICU & \multirow{2}{*}{\shortstack{Risk Prediction,\\ Diagnosis}} & \multirow{2}{*}{\shortstack{W-F1, R@k, \\ AUC, F1}} & MIMIC-III \\
 & & & & & Hospital & & & MIMIC-IV \\
\hline

%%%%%%%%%%%%%%%%%%%%%%%% 2025 %%%%%%%%%%%%%%%%%%%%%%%%%%%%%%%
\multicolumn{2}{c}{\textbf{2025 Year:}} \\
\hline
\cite{xiao2025xtsformer} & Feature & Type Only & $\checkmark$ & - & Hospital & Risk Prediction & ACC, F1, RMSE, NLL & Private \\
\hline
\cite{dai2025rotary} & Feature & Type Only & $\checkmark$ & - & ICU & \shortstack{Risk Prediction,\\ Diagnosis} & ACC, RMSE, LL & MIMIC-II \\
\hline
\cite{zhao2025balancing} & Feature & Type Only & $\checkmark$ & - & Hospital & Trajectory Generation & RMSE, Error Rate & Private \\
\hline
\cite{draxler2025transformers} & Feature & Type \& Value & $\checkmark$ & - & Hospital & Risk Prediction & NLL & EHRSHOT \\
\hline
\multirow{2}{*}{\cite{lee2025clinical}} & \multirow{2}{*}{Feature} & \multirow{2}{*}{Multimodal} & \multirow{2}{*}{$\checkmark$} & \multirow{2}{*}{-} & ED & \multirow{2}{*}{Risk Prediction} & \multirow{2}{*}{AUC, AUPRC, F1} & UCLA Health \\
& & & & & ICU & & & MIMIC-IV \\
\hline
\cite{seinen2025using} & Feature & Multimodal & $\checkmark$ & - & Primary Care & Clinical Coding & Jaccard & IPCI \\
\hline
\cite{thakur2025joint} & Feature & Multimodal & $\checkmark$ & - & Hospital & Risk Prediction & Sensitivity, Specificity & Private \\
\hline
\cite{dao2025curenet} & Feature & Multimodal & $\checkmark$ & - & ICU & Risk Prediction & ACC, F1, R@k, NDCG & MIMIC-III\\
\hline
\multirow{2}{*}{\cite{ma2025memorize}} & \multirow{2}{*}{Feature} & \multirow{2}{*}{Type Only} & \multirow{2}{*}{$\checkmark$} & \multirow{2}{*}{-} & \multirow{2}{*}{ICU} & \multirow{2}{*}{Risk Prediction} & F1, R@k, & MIMIC-III\\
& & & & & & & AUC, F1 & MIMIC-IV \\
\hline
\multirow{2}{*}{\cite{yin2025sepsiscalc}} & \multirow{2}{*}{Feature} & \multirow{2}{*}{Type \& Value} & \multirow{2}{*}{$\checkmark$} & \multirow{2}{*}{-} & \multirow{2}{*}{ICU} & \multirow{2}{*}{Risk Prediction} & \multirow{2}{*}{AUC, F1, Recall} & MIMIC-III \\
& & & & & & & & AmsterdamUMCdb \\
\hline
\cite{kauffman2025infehr} & Feature & Multimodal & $\checkmark$ & - & Hospital & Phenotyping & AUC, AUPRC & Private \\
\hline
\cite{wang2025efficient} & Feature & Type Only & $\checkmark$ & - & Primary Care & Survival Analysis & C-index & ORCPD \\
\hline
\cite{post2025harnessing} & Feature & Type Only & $\checkmark$ & - & Primary Care & Admission Prediction & AUC, AUPRC & SAIL \\
\hline
\multirow{2}{*}{\cite{song2025trajgpt}} & \multirow{2}{*}{Feature} & \multirow{2}{*}{Type Only} & \multirow{2}{*}{-} & \multirow{2}{*}{Next-tok} & Primary Care & \multirow{2}{*}{\shortstack{Risk Prediction, \\ Medication Rec.}} & \multirow{2}{*}{R@k, AUC} & PopHR \\
& & &  & & ICU & & & eICU \\
\hline
\cite{tariq2025adaptable} & Feature & Type \& Value & - & Next-tok & Hospital & Risk Prediction & AUC, SEN, SPE & MIMIC-IV \\
\hline
\multirow{2}{*}{\cite{theodorou2025improving}} & \multirow{2}{*}{Feature} & \multirow{2}{*}{Type \& Value} & \multirow{2}{*}{-} & \multirow{2}{*}{Next-tok} & \multirow{2}{*}{ICU} & \multirow{2}{*}{\shortstack{Trajectory Generation, \\ Risk Prediction}} & \multirow{2}{*}{F1, DI, TI} & MIMIC-IV \\
 & & & & & & & & eICU \\
\hline
\cite{shestov2025llm4es} & Event & Type \& Value & - & Next-tok & ICU & User Classification & AUC, MAE & MIMIC-IV \\
\hline
\cite{renc2025foundation} & Event & Type \& Value & - & Next-tok & ICU & Risk Prediction & \shortstack{AUC, AUPRC, \\ MAE} & MIMIC-IV \\
\hline
\cite{zhu2025medtpe} & Event & Type \& Value & - & Next-tok & ICU & Risk Prediction & F1 & MIMIC-IV \\
\hline

\cite{seinen2025using} & Feature & Multimodal & - & Skip-gram & Primary Care & Clinical Coding & S2T\%, T2S\% & IPCI Dutch GP \\
\hline
\cite{pradhan2025enhancing} & Feature & Type Only & - & Contrastive & Hospital & Clinical Coding & ACC, R@k & Private \\
\hline
\cite{amirahmadi2025trajectory} & Feature & Type Only & - & Contrastive & ICU & \shortstack{Risk Prediction,\\ LoS} & AUC, ACC & MIMIC-IV \\
\hline

%%%%%%%%%%%%%%%%%%%%%%%% 2026 %%%%%%%%%%%%%%%%%%%%%%%%%%%%%%%
\multicolumn{2}{c}{\textbf{2026 Year:}} \\
\hline
\multirow{2}{*}{\cite{chen2026cross}} & \multirow{2}{*}{Event} & \multirow{2}{*}{Type \& Value} & \multirow{2}{*}{$\checkmark$} & \multirow{2}{*}{-} & ICU & Risk Prediction & \multirow{2}{*}{AUC, AUPRC, F1} & MIMIC-IV \\
& & & & & ED, Hospital, ICU & Risk Prediction, LoS & & EHRSHOT \\
\hline
\cite{yin2026dual} & \multirow{2}{*}{Feature} & \multirow{2}{*}{Type \& Value} & \multirow{2}{*}{$\checkmark$} & \multirow{2}{*}{-} & \multirow{2}{*}{ICU} & \multirow{2}{*}{Risk Prediction} & \multirow{2}{*}{AUC, AUPRC} & MIMIC-III \\
& & & & & & & & MIMIC-IV \\
\hline

\end{supertabular}

    \begin{minipage}{1.0\linewidth}
    \tiny
    \textbf{Abbreviations} 
    Area Under Curve (AUC), Area Under Precision–Recall Curve (AUPRC), Precision–Recall AUC (AUPRC),
    Accuracy (ACC), Accuracy@k (ACC@k), Precision (PRE), Recall (REC), F1-score (F1),
    Weighted F1-score (W-F1),
    Sensitivity (SEN), Specificity (SPE), Positive Predictive Value (PPV), Negative Predictive Value (NPV),
    Mean Average Precision (mAP), Mean Average Precision@k (MAP@k),
    Precision@k (P@k), Recall@k (R@k), Precision@1 (P@1), Recall@1 (R@1), Recall@N (R@N),
    Weighted AUC (WAUC or W-AUC),
    Jaccard Similarity (JS), Average Recall (AR), Validity (VAL),
    Log-likelihood (LL), Root Mean Squared Error (RMSE), Error Rate (ER),
    Mean Absolute Error (MAE), Mean Squared Error (MSE),
    Concordance index (C-idx), Cohen's Kappa (Kappa), Median Absolute Deviation (MAD),
    Normalised Discounted Cumulative Gain (nDCG), nDCG@k (NDCG@k),
    Topic Coherence at top-10 (TC@10), Normalised Key–Query Matching at N (NKQM@N),
    Hamming Loss (HL), Half-Life Utility (HLU),
    Pearson correlation (r), Normalised Nearest Average Accuracy (NNAA),
    Hospital Readmission Rate (HRR),
    Medication Recommendation (Med Rec).
    \end{minipage}
\end{center}

\end{document}